\documentclass[pdflatex,sn-nature]{sn-jnl}

\usepackage{graphicx}
\usepackage{amsmath,amssymb,amsfonts}
\usepackage{booktabs}
\usepackage{placeins}

\hypersetup{colorlinks=true,allcolors=blue,bookmarksdepth=3,bookmarksnumbered=true}

\begin{document}
\addtocontents{toc}{\protect\setcounter{tocdepth}{-10}}

\title{Non-ergodic Area Scaling of Andreev Transport in Al/AlO$_x$/Al Josephson Junctions}

\author*[1]{\fnm{David J.} \sur{Perello}}\email{djperell@amazon.com}

\author[1]{\fnm{Marco} \sur{Coraiola}}

\author[1]{\fnm{Minje} \sur{Park}}

\author[1]{\fnm{Cliff} \sur{Chen}}

\author[1]{\fnm{Hanho} \sur{Lee}}

\author[1]{\fnm{Abhishek} \sur{Banerjee}}

\author[1]{\fnm{Gil} \sur{Refael}}

\author[1]{\fnm{Yuki} \sur{Sato}}

\affil*[1]{\orgname{Amazon Web Services}, \orgdiv{Center for Quantum Computing}, \orgaddress{\city{Pasadena}, \state{California}, \postcode{91106}, \country{USA}}}

\abstract{State-of-the-art superconducting qubits rely on Al/AlO$_x$/Al Josephson junctions (JJs) as a nonlinear element with an ideal sinusoidal current-phase dependence, with its inductance set by the junction area and oxidation conditions. This idealization holds only if the many conduction channels through the amorphous AlO$_x$ barrier transmit weakly and almost identically. Here, we measure the second-order subgap Andreev current, which sets the second Josephson harmonic, in Al/AlO$_x$/Al JJs. For junction sizes relevant to qubits, we find that this current is dominated by $\lesssim 300$ effective channels whose transparencies exceed the ideal tunneling limit ($\lesssim10^{-6}$) by orders of magnitude. For small and high-aspect-ratio junctions, increasing the junction size does not average out the effect of these high transparency channels but instead adds additional channels of even greater transparency. The trend inverts once both the width and length of the junction exceed a critical length scale, above which the effective transparency decreases with area. This non-ergodic scaling of the Andreev current is reproduced by simulations and attributed to non-self-averaging statistics of the transparency distribution, which arise from variations in oxide thickness. The effect is largely invisible to the normal-state conductance used to target the inductance and reveals unexpected geometric and size-related trade-offs for scaling superconducting qubit systems to large numbers of junctions.}


\pdfbookmark[0]{Main text}{maintext}
\maketitle

The Josephson junction is the cornerstone of superconducting quantum information processing, providing the nonlinearity essential for quantum operations~\cite{josephson1962possible, josephson1974discovery}. In particular, aluminum-oxide (AlO$_x$)-based tunnel Josephson junctions (JJs) have enabled superconducting hardware to become a leading platform for fault-tolerant quantum computing~\cite{google2021exponential, sivak2023real, krinner2022realizing, putterman2025hardware}, and form crucial elements for quantum-limited amplifiers~\cite{roy2016introduction},microwave photon detectors~\cite{albertinale2021detecting} and SQUID magnetometers~\cite{fagaly2006squid}. As quantum-processor architectures scale toward $10^5$ junctions~\cite{Kim2023,Morvan2024,Bravyi2022} and ground-plane materials improve~\cite{place2021new, ganjam2024surpassing,msecTa2025}, the junction itself becomes a limiting source of device variability and decoherence~\cite{msecTa2025,Willsch2024}.

The supercurrent in JJs is carried by $N_\mathrm{ch}$ conducting channels, each of which forms an Andreev bound state (ABS). Summing over the $N_\mathrm{ch}$ contributions in the zero-temperature limit $k_B T\ll\Delta$ appropriate to our measurements, we obtain the junction current-phase relation (CPR):
\begin{equation}
I(\phi) = \frac{e\Delta^2}{2\hbar}\sin\phi \sum_{i=1}^{N_\mathrm{ch}} \frac{\tau_i}{E_i^A(\phi, \tau_i)}  = \sum_{m=1}^{\infty} a_m\sin(m\phi),
\label{eq:1}
\end{equation}

where $\pm E_i^A = \pm\Delta\sqrt{1 - \tau_i\sin^2(\phi/2)}$ are the energies of the $i^\mathrm{th}$ ABS pair~\cite{Beenakker1991,RevModPhys.76.411}, $\phi$ is the superconducting order parameter phase difference, and $\tau_i$ is the $i^\mathrm{th}$ channel transparency. The final equality is obtained through a Fourier series expansion and emphasizes the presence of higher harmonics within a general JJ. A measure of junction non-ideality is the CPR skew, given by the ratio of the second to first harmonics $|a_2/a_1|$, which results from the presence of {\it leaky} channels with large transparencies. Non-zero skew causes unpredictable deviations in transmon excited states~\cite{Willsch2024}, reduces readout fidelity through measurement-induced transitions~\cite{mist-cpr}, and increases dephasing from trapped quasiparticles and critical-current fluctuations~\cite{PhysRevB.86.184514}.

\begin{figure}[t]
\centering

\includegraphics[width=0.5\linewidth]{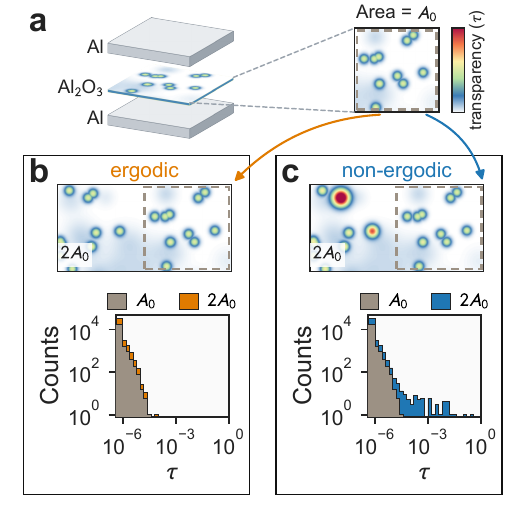}

\caption{Ergodic (self-averaging) versus non-ergodic (non-self-averaging)  area scaling of the tunneling transparency (model schematic). \textbf{a}, Exploded view of one Al/AlO$_x$/Al junction of area $A_\mathrm{jj}=A_0$ and the transparency map of its amorphous barrier (colour: local transparency $\tau$, white low to red high). \textbf{b},\textbf{c}, shows how a slice of the sample twice in size would look, under the two hypotheses. Below each map, the histogram of channel transparencies of the $A_0$ region alone (grey, identical in \textbf{b} and \textbf{c}) and of the full $2A_0$ junction (colour). Self-averaging, (\textbf{b}) implies that any sub-area is statistically representative of the whole so both $\langle\tau\rangle$ and $\langle\tau^2\rangle$ are unchanged. In contrast, non self averaging sample, (\textbf{c}) the larger area samples even deeper into the right tail of the fat-tailed distribution, and adds channels of increasing transparency. These are the rare events responsible for the non-self-averaging behavior; $\langle\tau\rangle$ barely changes while $\langle\tau^2\rangle$ increases with area.}\label{fig:Fig1}
\end{figure}

For a junction obeying Eq.~\ref{eq:1}, expanding to second order in $\tau$ and summing over channels gives $|a_2/a_1| = \langle\tau^2\rangle/(8\langle\tau\rangle)$. Since the Josephson energy $E_{Jm}$ of a higher harmonic is proportional to $a_m/m$, a ratio of $|E_{J2}/E_{J1}| = \langle\tau^2\rangle/(16\langle\tau\rangle)$ is expected purely from the CPR skew. Previous work extracted higher harmonic content from the transition spectra of transmon qubits~\cite{Willsch2024, kim2025emergentharmonicsjosephsontunnel}. To disentangle the $|E_{J2}/E_{J1}|$ contributions intrinsic to the junction from those of the circuit inductance, the former was assumed to be independent of junction area $A_\mathrm{jj}$~\cite{kim2025emergentharmonicsjosephsontunnel} and thus distinguishable from the inductive contribution, which is proportional to $A_\mathrm{jj}$. Fundamentally, this assumption relies on the spatial averages of $\tau$ and $\tau^2$ converging to their ensemble values independently of junction area. That is, the ratio $\langle\tau^2\rangle/\langle\tau\rangle$ is constant and area-independent, and the junction scales \textit{ergodically}: its size does not influence the nature of first- and second-order Andreev transport. Figure~\ref{fig:Fig1} illustrates this concept by comparing the tunneling transparency map of one small junction of area $A_0$ with that of a junction of twice the area, $2A_0$. With ergodic scaling, the larger junction has effectively the same transparency distribution and reproduces the hot-spot (high-transparency) statistics of the smaller junction; both $\langle\tau\rangle$ and $\langle\tau^2\rangle$ are essentially unchanged, because the smaller region is statistically representative of the larger area. In the non-ergodic case, rare and extreme high-transparency channels become more common as the area grows. The tunneling transparency statistics are then non-self-averaging~\cite{gilrefael2004}: the most transparent channels disproportionately affect $\langle\tau^2\rangle \propto E_{J2}$, resulting in a strong area dependence. Conversely, $\langle\tau\rangle \propto G_n$, where $G_n$ is the normal-state conductance of the JJ, is essentially unchanged. This implies that the room-temperature conductance measurements used in frequency targeting are insensitive to this area scaling, making it challenging to discriminate from inductance-related effects in transmon experiments.

Here, we characterize the underlying transparencies and junction scaling by directly probing the second-order Andreev current using DC subgap transport measurements of 134 JJs with varied area ($A_\mathrm{jj}$), oxide thickness ($d$), and aspect ratio ($w/l$, where $w$, $l$ are the junction width and length). The large and diverse sample set allows observation of unexpected scaling of the second-order Andreev current, which depends on both the junction area and aspect ratio. For small junctions and junctions with large aspect ratio, the dependence is consistent with the non-ergodic scaling in Fig.~\ref{fig:Fig1}. We find that an emergent threshold length scale $l_\mathrm{th}$ exists whereby junctions with both $l,w>l_\mathrm{th}$ display improved oxide thickness uniformity, reduced effective transparency, and behavior closer to the ideal junction with a sinusoidal CPR. For junction sizes typically relevant to transmon qubits, a small number $\lesssim 300$ of rare high transparency channels dominate the contribution to $\langle\tau^2\rangle$ and therefore $E_{J2}$. Our results provide an alternative interpretation of recent area-dependent $E_{J2}/E_{J1}$ observations, and demonstrate how junction microscopics and the underlying distribution of transparencies constrain the design of scalable qubit architectures.

The 134 Al/AlO$_x$/Al JJs studied here are fabricated following standard shadow-evaporation processes, with bottom electrode width $w$ and top electrode width $l$ ranging from $0.1~\mu$m up to $2.0~\mu$m, aspect ratios $0.25<w/l<10$, and oxidation exposures $\mathcal{E}$, the product of the static O$_2$ pressure and the oxidation time, from 4 to 480~Torr$\cdot$min. Junction areas $A_\mathrm{jj} = l\,w + 2\,l\,t_\mathrm{b}$ include the overlap of the top electrode with the two sidewalls of the bottom electrode of thickness $t_\mathrm{b}$, while $l$ and $w$ are SEM-measured for Square Set A and design values otherwise (Supplementary Section~\ref{sec:S1}). The junctions with $w/l\ne1$ comprise three different fabrication batches and are referred to as high-aspect-ratio (HAR) Sets A, B, and C. Sets A and B each contain devices with a fixed short dimension ($l = 200$ or $300$~nm), while Set C varies both $l$ and $w$. For square devices, set A was fabricated in a single deposition step while set B underwent multi-step fabrication. DC measurements are taken at 10~mK in a Bluefors LD dilution refrigerator, with RC- and powder-filtered lines (Fig.~\ref{fig:FigS1}).

\begin{figure}[t]
\centering
\includegraphics[width=0.5\linewidth]{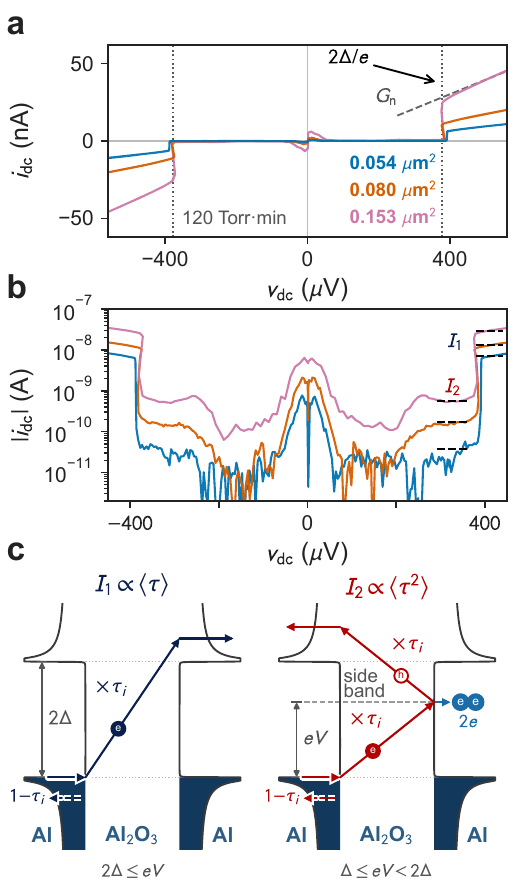}
\caption{DC transport characteristics of Al/AlO$_x$/Al Josephson junctions.
\textbf{a},~Signed $i_\mathrm{dc}$--$v_\mathrm{dc}$ characteristics for three junctions with $A_\mathrm{jj} = 0.054$, $0.080$, and $0.153~\mu\mathrm{m}^2$ at a fixed oxidation exposure of 120~Torr$\cdot$min. Dotted vertical lines mark $v_\mathrm{dc}=\pm2\Delta/e$; the dashed guide indicates the normal-state conductance $G_n$.
\textbf{b},~$|i_\mathrm{dc}|$ versus $v_\mathrm{dc}$ on a logarithmic scale for the same three junctions as in \textbf{a}. Dashed lines denote the extraction of the step amplitudes $I_1$ and $I_2$.
\textbf{c},~Schematic of the two lowest-order Andreev tunneling processes. For $eV>2\Delta$ a single quasiparticle excitation and barrier traversal occurs, giving $I_1\propto\langle\tau\rangle$. For $\Delta<eV<2\Delta$ transport requires coherent crossings of a hole-like (left-traveling) and electron-like quasiparticle (right-traveling), resulting in an additional cooper pair ($2e$) on the right electrode. This process has amplitude $I_2\propto\langle\tau^2\rangle$, with each coherent channel crossing $i$ contributing a factor $\tau_i$.}\label{fig:Fig2}
\end{figure}

\begin{figure}[p]
\centering
\includegraphics[width=1\linewidth]{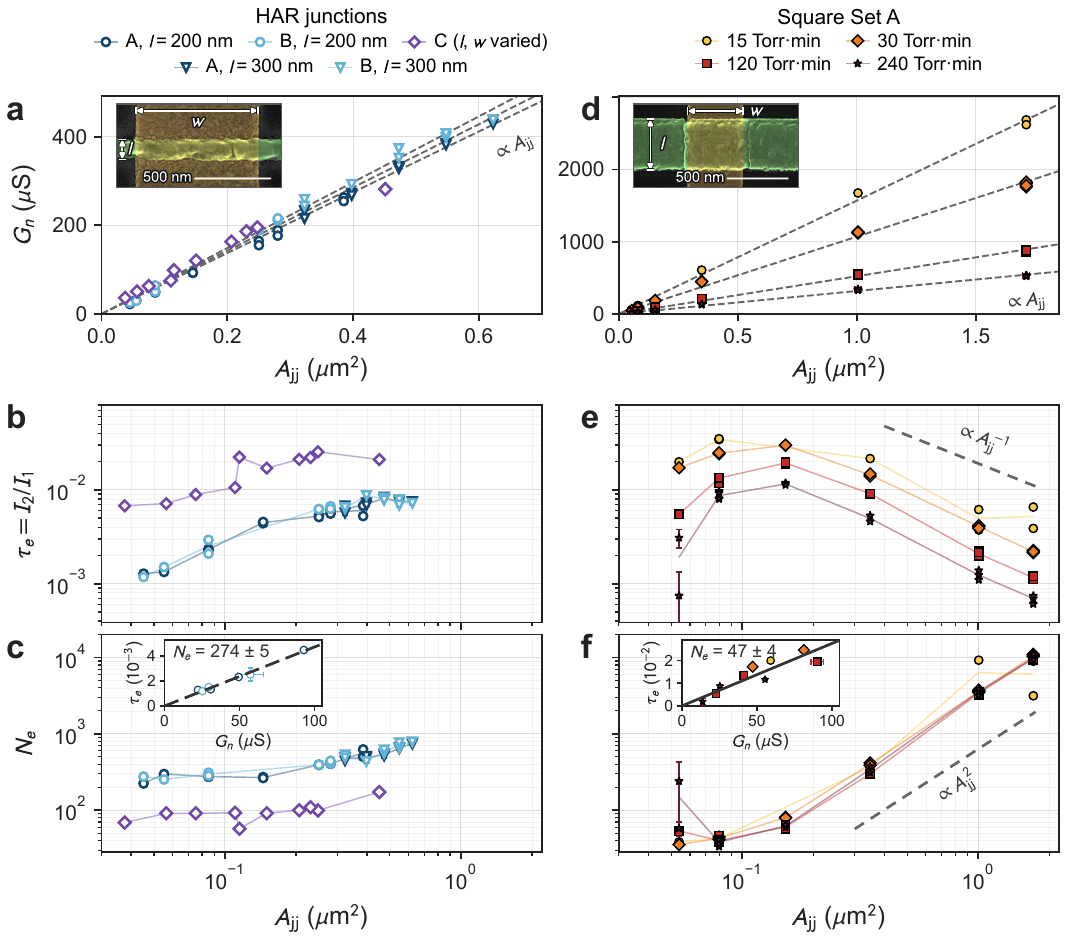}
\caption{Normal-state conductance, effective transparency ($\tau_e = I_2/I_1$) and effective channel number ($N_e$) versus junction area for the high-aspect-ratio (HAR) sets (\textbf{a}-\textbf{c}) and Square Set A (\textbf{d}-\textbf{f}).
\textbf{a},\textbf{d},~$G_n$ versus $A_\mathrm{jj}$. HAR Sets A and B have a short dimension of either $l = 200$~nm or $l = 300$~nm and correspond to different fabrication batches while set C varies both $l$ and $w$. Square Set A comprises four oxidation exposures (15, 30, 120, and 240~Torr$\cdot$min). Dashed lines are one-parameter fits $G_n = C\,A_\mathrm{jj}$ ($\propto A_\mathrm{jj}$) to each series. Insets: false-color SEM images of a representative HAR and square junction with the electrode widths $l$, $w$ marked (scale bars 500~nm).
\textbf{b},\textbf{e},~$\tau_e$ versus $A_\mathrm{jj}$. $\tau_e$ was previously assumed to be constant with junction area~\cite{kim2025emergentharmonicsjosephsontunnel}. Here it increases with area in every HAR set, whereas the square junctions are non-monotonic with peak near $A_\mathrm{jj} \approx 0.15~\mu\mathrm{m}^2$. Large junctions show decreasing $\tau_e$ as $\propto A_\mathrm{jj}^{-1}$ (dashed guide).
\textbf{c},\textbf{f},~Point-wise $N_e$ (Eq.~\ref{eq:3}) versus $A_\mathrm{jj}$. The HAR samples show a strongly sublinear $N_e$, while square devices have constant $N_e \approx 50$ for $A_\mathrm{jj}<0.15~\mu\mathrm{m}^2$ and then $N_e = (3.3\pm0.2)\times10^{3}\,A_\mathrm{jj}^{2.07\pm0.07}$ ($A_\mathrm{jj}$ in $\mu\mathrm{m}^2$) for larger junctions ($\propto A_\mathrm{jj}^{2}$, dashed guide). Insets: $\tau_e$ versus $G_n$ for junctions with $G_n<100~\mu\mathrm{S}$. Independent of oxidation condition, the points are collinear with inverse slope corresponding to a constant $N_e = 274\pm 5$ (fit to HAR Sets A and B) and $N_e = 47\pm 4$ (Square Set A, solid).
}\label{fig:Fig3}
\end{figure}

The resulting voltage-driven $i_\mathrm{dc}$--$v_\mathrm{dc}$ measurements (Fig.~\ref{fig:Fig2}a) show behavior typical of tunnel JJs, with a supercurrent peak at $v_\mathrm{dc}=0$ and the onset of a quasiparticle current for $v_\mathrm{dc}\ge2\Delta/e$, where $\Delta = 180~\mu$eV is the Al order parameter. Figure~\ref{fig:Fig3}a,d plot $G_n$ as a function of $A_\mathrm{jj}$ for the rectangular junction sets and for square junctions at four oxidation conditions. In every set the normal-state conductance $G_n$ increases linearly with $A_\mathrm{jj}$, and it decreases with oxidation exposure according to a power-law dependence (Fig.~\ref{fig:FigS3}). When plotted on a logarithmic scale (Fig.~\ref{fig:Fig2}b), a clear $i_\mathrm{dc}$ step becomes visible near $v_\mathrm{dc}=\Delta/e$. The current steps $I_1$ and $I_2$ (dashed lines in Fig.~\ref{fig:Fig2}b) result from sideband scattering of quasiparticles mediated by ABSs, a process which is represented schematically in Fig.~\ref{fig:Fig2}c. For first order scattering $n=1$, a quasiparticle is excited by an applied voltage $v_\mathrm{dc}\ge 2\Delta/e$, crossing the insulating barrier with transmission probability $\tau_i$ ($i \in \{1,2,\ldots,N_\mathrm{ch}\}$ conduction channels). For $n=2$, a right-moving quasiparticle is excited by $\Delta/e \leq v_\mathrm{dc} < 2\Delta/e$, scattering into a sideband channel on the opposite electrode with probability $\tau_i$ while, coherently, a hole is excited and scattered with probability $\tau_i$ into the $n=2$ sideband above the gap edge, leaving one cooper pair on the right electrode. Thus, the $n$ coherent excitations on the ladder result in current step $I_n\propto \tau_i^n$~\cite{Bratus1997,RevModPhys.76.411,resonantandreev,PhysRevLett.73.2611,PhysRevB.27.6739,Cuevas1996,Flensberg1988,Beenakker1991}. In practice, only the first- ($I_1$) and second-order ($I_2$) current steps are visible in our tunneling junctions due to the relatively small step size.

Having established the transport mechanism, the two primary observables reflecting the behavior of $\langle \tau \rangle,\langle \tau^2 \rangle$ are the effective number of channels $N_e$ and the effective transparency $\tau_e$, which are defined as:

\begin{equation}
\tau_e \equiv \frac{I_{2}}{I_{1}}=\frac{\langle\tau^2\rangle}{\langle\tau\rangle}
\end{equation}

\begin{equation}
N_e = \frac{hG_{n}}{2e^2}\frac{I_{1}}{I_{2}}.
\label{eq:3}
\end{equation}

The normal-state conductance $G_n$ is extracted from a linear fit to the ohmic branch well above $2\Delta/e$, and $I_1,I_2$ are extracted as in Fig.~\ref{fig:Fig2}b. The values of $\tau_e$ and $N_e$ from experimental $i_\mathrm{dc}$-$v_\mathrm{dc}$ data are plotted in Fig.~\ref{fig:Fig3}, with additional datasets available in the Supplementary Information (Sec.~\ref{sec:S3}). Due to the starkly differing area dependencies, we segregate the data from rectangular junctions with $w/l\ne1$ (Fig.~\ref{fig:Fig3}a-c) and square junctions with $w/l=1$ (Fig.~\ref{fig:Fig3}d-f). Baseline $G_n$ vs $A_{jj}$ curves in Fig.~\ref{fig:Fig3}a,c, show typical linear behavior and it is only by looking at the subgap that the scaling behavior begins to differ. Focusing on HAR in Fig.~\ref{fig:Fig3}b, we first observe that the $\tau_e$ data from HAR C are co-linear, even though the set spans junctions with a range of top- and bottom-electrode widths and lengths. More strikingly, however, is that $\tau_e$ depends on $A_\mathrm{jj}$- increasing by an order of magnitude across the area range studied in all HAR datasets. The square junctions in Fig.~\ref{fig:Fig3}e also show an unexpected, area-dependent $\tau_e$, but here the trend is non-monotonic. An emergent threshold length scale $l_\mathrm{th}$ appears such that for $l,w>l_\mathrm{th}$, $\tau_e$ peaks and begins to \textit{decrease} as $A_\mathrm{jj}^{-1}$. We observe that $l_\mathrm{th}$ is process dependent, such that Square Set A inverts at $A_\mathrm{jj}\approx0.14~\mu\mathrm{m}^2$ ($l_\mathrm{th}\approx0.34~\mu$m), while square junctions on a Set B inverts near $A_\mathrm{jj}\approx0.4$--$0.5~\mu\mathrm{m}^2$ ($l_\mathrm{th}\approx0.67~\mu$m) (Fig.~\ref{fig:FigS2}). Within the sets of devices, the value of $l_\mathrm{th}$ is also independent of the oxidation exposure.

The source of this area dependence can be traced to the dependence of $I_2$ on junction area, since all junction sets have $G_n\propto I_1\propto A_\mathrm{jj}$ (Fig.~\ref{fig:Fig3}a,d). The inversion of $\tau_e$ in square junctions results from $I_2$ saturating at $A_\mathrm{jj}=A_\mathrm{th}$, the area of a junction of side $l_\mathrm{th}$ (Eq.~\ref{eq:4}), to a constant value as $A_\mathrm{jj}$ increases (Fig.~\ref{fig:FigS4}) while $I_1$ continues to increase thus lowering $\tau_e$. We also note that the size of the edge-overlap between the top and bottom electrodes (``rising edge") does not affect $\tau_e$ appreciably, since HAR A, B, and C all display a similar trend.

Next, Using Eq.~\ref{eq:3}, $N_e$ is plotted as a function of $A_{jj}$ in Fig.~\ref{fig:Fig3}c,f. $N_e$ is flat or increases sub-linearly for $A_\mathrm{jj}<A_\mathrm{th}$, and is independent of oxidation exposure for devices within a given fabrication batch. Fitting $\log(N_e)$ versus $\log(A_\mathrm{jj})$ for square junctions with $A_\mathrm{jj}>A_\mathrm{th}$, we find $N_e = (3.3\pm0.2)\times10^{3}\,A_\mathrm{jj}^{2.07\pm0.07}$ with $A_\mathrm{jj}$ in $\mu\mathrm{m}^2$. For comparison, in a homogeneous tunnel barrier $N_e$ should scale linearly with area, up to $N_e=N_\mathrm{ch}$ for an ideal junction. Zooming in on the small area region, we plot $\tau_e$ directly as a function of $G_n$ (insets of Fig.~\ref{fig:Fig3}c,f). Data points from the small junctions ($l,w < l_\mathrm{th}$) fabricated in the same lot collapse to a common exposure-invariant line when $G_n< 100~\mu\mathrm{S}$. The inverse slope gives constant $N_e = 47\pm 4$ and $N_e = 274\pm 5$ for the Square Set A and HAR junctions, respectively. This conductance range corresponds to $E_J/h < 14~\mathrm{GHz}$ (considering $\Delta = 180~\mu\mathrm{eV}$), which overlaps with typical $E_J$ in transmon circuits. A similar analysis was used previously in small break junctions with few channels, but colinearity is unexpected in planar SIS tunnel junctions with $\tau_e\ll1$~\cite{PhysRevLett.73.2611}.

The experimental data above indicates that small square devices with $l,w<l_\mathrm{th}$ and all HAR devices have increasing $\tau_e$ and effectively constant $N_e$. This behavior is consistent with increased area resulting in the addition of channels with increasing transparency that dominate the transport. Conversely, junctions with both $l,w>l_\mathrm{th}$ have $\tau_e\propto A_\mathrm{jj}^{-1}$ and $N_e\propto A_\mathrm{jj}^2$. Since $I_2$ is constant or slowly varying for $l,w>l_\mathrm{th}$ (Fig.~\ref{fig:FigS4}), we infer that only low transparency channels are added with increased area.

\begin{figure}[p]
    \centering
    \includegraphics[width=0.97\linewidth]{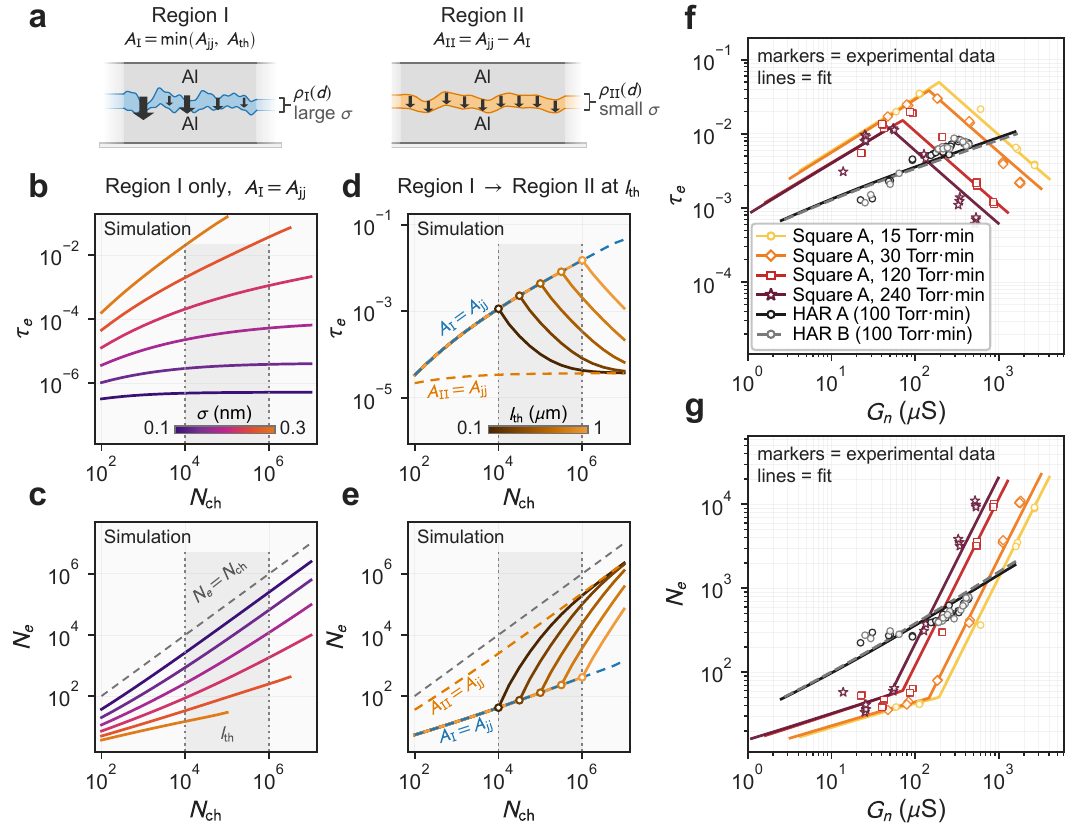}
    \caption{Two-region disorder model and fits to the experimental data.
    \textbf{a},~Schematic. Region I is a disordered oxide with thickness distribution $\rho_\mathrm{I}(d)$ (large $\sigma$) occupying $A_\mathrm{I}=\min(A_\mathrm{jj},A_\mathrm{th})$: the whole junction when $\min(l,w)<l_\mathrm{th}$ and a fixed area $A_\mathrm{th}$ otherwise (Eq.~\ref{eq:4}). Region II is a more uniform oxide, $\rho_\mathrm{II}(d)$ (small $\sigma$), covering the remaining area $A_\mathrm{II}=A_\mathrm{jj}-A_\mathrm{I}$. Arrows indicate the transparency of the channels within each region, larger arrows corresponding to higher transparencies.
    \textbf{b},\textbf{c},~Simulated $\tau_e$ and $N_e$ versus channel number $N_\mathrm{ch}=n_\mathrm{ch}A_\mathrm{jj}$ for a junction composed only of Region I ($A_\mathrm{I}=A_\mathrm{jj}$), for $\sigma_\mathrm{I} = 0.10$, 0.14, 0.18, 0.22, 0.26, and 0.30~nm (color bar) at fixed $d_\mathrm{I}=1.40$~nm. For small $\sigma_\mathrm{I}$, $\tau_e$ is size-independent and $N_e\propto N_\mathrm{ch}$ (gray dashed line, $N_e=N_\mathrm{ch}$); increasing $\sigma_\mathrm{I}$ makes both observables strongly size-dependent. Curves are terminated where the thinnest channel of the distribution reaches $\tau=1$. The shaded band marks $N_\mathrm{ch}=n_\mathrm{ch}l_\mathrm{th}^2$ for $l_\mathrm{th}=0.1$-$1~\mu$m, the range swept in \textbf{d},\textbf{e}.
    \textbf{d},\textbf{e},~Region I~$\to$~Region II transition, sweeping $l_\mathrm{th}$ from 0.1 to 1~$\mu$m (color bar, shaded band) at fixed $\rho_\mathrm{I}$ with $(d_\mathrm{I},\sigma_\mathrm{I})=(1.40,0.25)$~nm and $\rho_\mathrm{II}$ with $(d_\mathrm{II},\sigma_\mathrm{II})=(1.05,0.10)$~nm. Open circles mark $N_\mathrm{ch}=n_\mathrm{ch}l_\mathrm{th}^2$, where the Region-I channel count saturates (the simulation carries no sidewall term); dashed lines are the pure Region-I ($A_\mathrm{I}=A_\mathrm{jj}$, blue) and pure Region-II ($A_\mathrm{II}=A_\mathrm{jj}$, orange) limits.
    \textbf{f},\textbf{g},~Experimental $\tau_e$ and $N_e$ versus $G_n$ (markers) with model fits (lines) for the four Square Set A exposure series and HAR Sets A and B (100~Torr$\cdot$min). The extracted parameters are listed in Table~\ref{tab:Table1}. All simulations use $\theta=1.15$~eV, $n_\mathrm{ch}=10^6~\mu\mathrm{m}^{-2}$, and deterministic order statistics.
    }
    \label{fig:Fig4}
\end{figure}

Based on these trends, we introduce an analysis framework which utilizes the area-dependence to extract and characterize the underlying channel transparency distribution $\rho(\tau)$. The data suggests two distinct disorder regions in the junctions, denoted Region I and Region II and shown schematically in Fig.~\ref{fig:Fig4}a. Region I corresponds to the small square devices and HAR junctions. Its oxide is assumed to have greater spatial variability, with mean thickness $d_\mathrm{I}$ and thickness standard deviation $\sigma_\mathrm{I}$. Region II emerges in larger square junctions and has more uniform thickness, resembling an ideal oxide. The oxide thickness distribution in this region has mean $d_\mathrm{II}$ and standard deviation $\sigma_\mathrm{II}$, with $\sigma_\mathrm{II} < \sigma_\mathrm{I}$. The observed constant $I_2$ (Fig.~\ref{fig:FigS4}), $\tau_e\propto A_\mathrm{jj}^{-1}$ and $N_e\propto A_\mathrm{jj}^2$ therefore reflect the crossover from Region I to Region II. The disordered Region I remains fixed in size, while only low-disorder Region II is added for $l,w>l_\mathrm{th}$ at a rate proportional to $A_\mathrm{jj}$. Algebraically, the areas of Region I (Region II), $A_\mathrm{I}$ ($A_\mathrm{II}$), are then defined as

\begin{equation}
\begin{aligned}
A_\mathrm{I} &=
\begin{cases}
A_\mathrm{jj}, & \min(l,w) < l_\mathrm{th}\\[2pt]
A_\mathrm{th} \equiv l_\mathrm{th}^{2} + 2\,l_\mathrm{th}\,t_\mathrm{b}, & \min(l,w) \ge l_\mathrm{th}
\end{cases}
\qquad
A_\mathrm{II} = A_\mathrm{jj} - A_\mathrm{I},
\end{aligned}
\label{eq:4}
\end{equation}
where $A_\mathrm{th}$ is the area, sidewalls included, of a square junction of side $l_\mathrm{th}$.

We model the tunneling behavior in Regions I and II assuming spatially inhomogeneous direct tunneling across a trapezoidal energy barrier shape, hence applying the Simmons model at $v_\mathrm{dc}\approx0$~\cite{improvementsimmonsmodeltunnel,Simmons}. The low-voltage transparency at zero temperature is given by
\begin{equation}
\tau(d)=\frac{C}{d}\,e^{-2\kappa d},
\label{eq:5}
\end{equation}
where $d$ is the barrier thickness, $\kappa \equiv \sqrt{2 m_e \theta}/\hbar = 5.12~\mathrm{nm}^{-1}\mathrm{eV}^{-1/2}\,\sqrt{\theta}$, $m_e$ is the electron mass, $\theta$ is the potential energy barrier height, $C=\sqrt{2m_e\theta}/(2h\, n_\mathrm{ch})$, and $n_\mathrm{ch}=10^6~\mu\mathrm{m}^{-2}$ is the areal channel density assumed for this simulation. The number of simulated conducting channels is then given as $N_\mathrm{ch}=n_\mathrm{ch}A_\mathrm{jj}$. In this model, the tunneling transmission $e^{-2\kappa d}/d$ acts as a sieve whereby the channels located in regions with the thinnest  barrier have an over-sized influence on $\langle\tau\rangle$, compounded at second order in $I_2\propto\langle \tau^2 \rangle$. Choosing an appropriate thickness distribution $\rho(d)$, is therefore critical to the model accuracy, and we show in Sec.~\ref{sec:S4} that a normal $\rho(d)$ more accurately reproduces the observed experimental results. Refer to Sec.~\ref{sec:S4} for comparisons of functional forms of $\rho(d)$ and any details of the simulations and fitting methodology.

To connect the model to the observed trends, we begin with simulations of $N_e$ and $\tau_e$ for representative parameters (Fig.~\ref{fig:Fig4}b-e). For the single-region case (HAR devices and small-area square junctions, which contain only Region I), we fix $d_\mathrm{I} = 1.40$~nm and $\theta = 1.15$~eV and vary $\sigma_\mathrm{I}$. Within a proper parameter range, an area dependence in $\tau_e$ is reproduced for $\kappa\sigma \gtrsim 0.8$ (and strongly so for $\kappa\sigma \gtrsim 1$), which corresponds to $\sigma \gtrsim 0.15$~nm ($\gtrsim 0.18$~nm) at $\theta=1.15$~eV in Fig.~\ref{fig:Fig4}b,c. Likewise, $N_e$ is linear with $N_\mathrm{ch}$, but flattens out as $\sigma_\mathrm{I}$ is increased. Next, we simulate the behavior of large square junctions by adding a Region II with much lower disorder. Extending the results from Fig.~\ref{fig:Fig4}b,c, we fix parameters $d_\mathrm{I}=1.40$~nm, $\sigma_\mathrm{I}=0.25$~nm, $d_\mathrm{II} = 1.05$~nm and $\sigma_\mathrm{II}=0.10$~nm. The threshold length $l_\mathrm{th}$ is used as the floating parameter with results shown in Fig.~\ref{fig:Fig4}d,e. Both $\tau_e,N_e$ transition from the pure Region-I curve ($A_\mathrm{jj}=A_\mathrm{I}$) toward the pure Region-II curve ($A_\mathrm{jj}=A_\mathrm{II}$). In Region II, where $\sigma_\mathrm{II} < 0.2$~nm, $\tau_e$ decreases and $N_e$ eventually becomes linear in $N_\mathrm{ch}$ similar to the experimental data. Interestingly, this area dependence arises intrinsically from the sampling of the transparency distribution $\rho(\tau)$. Because $\rho(\tau)$ is fat-tailed, i.e., its right tail does not quickly decay to zero as $\tau$ increases, sampling more channels (or, equivalently, increasing the junction area) preferably picks up high-$\tau$ outliers, producing a scaling similar to that shown in Fig.~\ref{fig:Fig1}c. $\tau_e$ is therefore governed by non-self-averaging behavior: its expectation value does not reflect the bulk of the channels, but is instead determined by rare, high-transparency outliers.

Assuming HAR junctions are pure Region I, experimental data is fit using a cost-minimization algorithm as described in the Supplementary Information, Section~\ref{sec:S5}. For the square junctions, the two smallest areas are assumed to be Region-I dominant (for the 240~Torr$\cdot$min series the second- and third-smallest, the smallest being an outlier), thus parameters $(d_\mathrm{I},\sigma_\mathrm{I})$ are first fit per oxidation exposure using data from just these two areas. The first-stage result then seeds a joint fit over the entire area range, which determines $l_\mathrm{th}$ and $d_\mathrm{II}$. Since the measured $\tau_e$ keeps decreasing and does not saturate at large $A_\mathrm{jj}$ (Fig.~\ref{fig:Fig4}f), $\sigma_\mathrm{II}$ can only be bounded from above, $\sigma_\mathrm{II} \lesssim 0.14$~nm. The bound is set by the Region-II $\tau_e$ versus $A_\mathrm{jj}$ curves, which must satisfy $\tau_e(A_\mathrm{jj}) > \tau_e(A_\mathrm{II})$ at every area. All fits below this bound are equivalent (Fig.~\ref{fig:FigS9}), therefore $\sigma_\mathrm{II}$ is fixed at $0.10$~nm during the optimization. The fit results are shown in Fig.~\ref{fig:Fig4}f,g and the extracted parameters are listed in Table~\ref{tab:Table1}. The model accurately reproduces the trends in both $\tau_e, N_e$ as function of $A_\mathrm{jj},G_n$, including the sharp change of trends at $l_\mathrm{th}$.

\begin{table}[!htb]
\centering
\small
\setlength{\tabcolsep}{4pt}
\caption{Region-I and Region-II fit parameters at $\theta = 1.15$~eV on the total junction area $A_\mathrm{jj} = l\,w + 2\,l\,t_\mathrm{b}$. $\mathcal{E}$ is the oxidation exposure (static O$_2$ pressure $\times$ oxidation time); $R_\mathrm{n} A_\mathrm{jj}$ is the median normal-state resistance$\cdot$area product of the junctions in each series (Supplementary Table~\ref{tab:TableS1}). The joint Square Set A fit gives $l_\mathrm{th} = 340$~nm ($A_\mathrm{th} = 0.141~\mu\mathrm{m}^2$); $\sigma_\mathrm{II} = 0.10$~nm.}
\label{tab:Table1}
\begin{tabular}{lcccccc}
\toprule
Device & $\mathcal{E}$ (Torr$\cdot$min) & $R_\mathrm{n} A_\mathrm{jj}$ ($\Omega\,\mu\mathrm{m}^2$) & $d_\mathrm{I}$ (nm) & $\sigma_\mathrm{I}$ (nm) & $\sigma_\mathrm{I} / d_\mathrm{I}$ & $d_\mathrm{II}$ (nm) \\
\midrule
Square A & 15 & 676 & 1.411 & 0.280 & 0.199 & 0.974 \\
Square A & 30 & 918 & 1.444 & 0.284 & 0.197 & 0.993 \\
Square A & 120 & 1906 & 1.493 & 0.283 & 0.189 & 1.070 \\
Square A & 240 & 3090 & 1.545 & 0.290 & 0.188 & 1.091 \\
\midrule
HAR A & 100 & 1495 & 1.306 & 0.225 & 0.173 & --- \\
HAR B & 100 & 1333 & 1.292 & 0.222 & 0.172 & --- \\
HAR C ($l,w$ var.) & 96 & 1243 & 1.388 & 0.261 & 0.188 & --- \\
\bottomrule
\end{tabular}
\end{table}

\begin{figure}[t]
\centering
\includegraphics[width=1\linewidth]{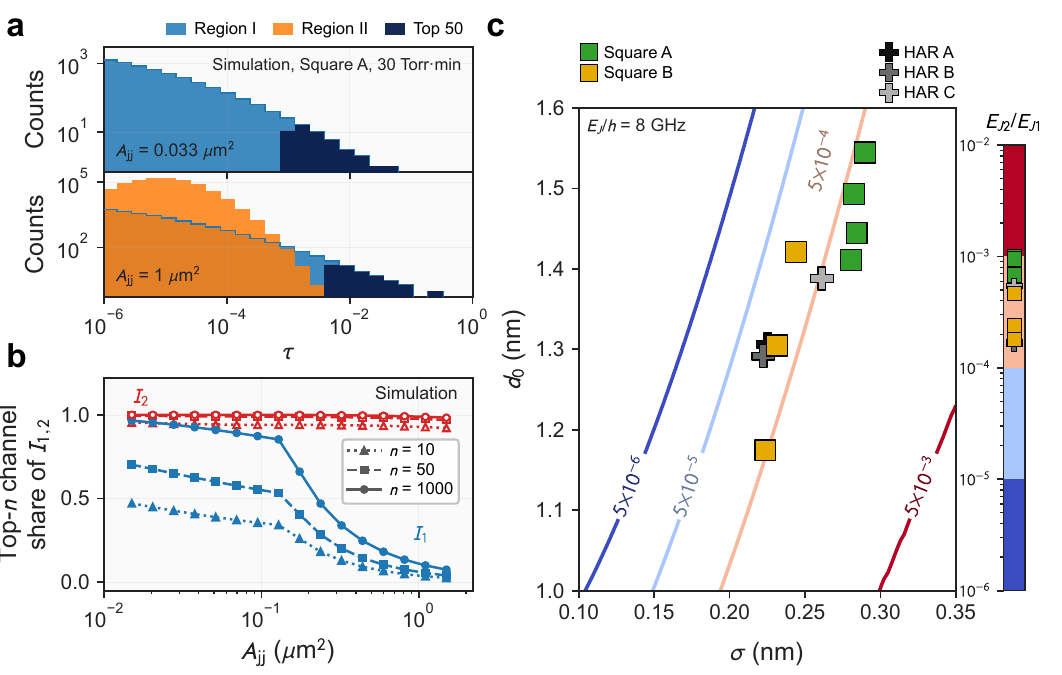}
    \caption{Simulated channel statistics and implications for qubit junctions.
    \textbf{a},~Histograms of channel transparencies $\tau_i$ simulated with the fitted Square Set A 30~Torr$\cdot$min parameters (Table~\ref{tab:Table1}) for devices of size $A_\mathrm{jj}=0.033$ (top) and $1.0~\mu\mathrm{m}^2$ (bottom). For both areas, all 50 of the most transparent channels (dark navy) are found in Region I, despite the large device being $\sim$86\% Region II channels.
    \textbf{b},~Fraction of $I_1$ and $I_2$ carried by the top $n=10$, 50, and 1000 most-transparent channels versus $A_\mathrm{jj}$ for the 30~Torr$\cdot$min simulation. The top 10 channels already carry $\gtrsim$90\% of $I_2$ at every area, while their share of $I_1$ decreases rapidly with area.
    \textbf{c},~Josephson-harmonics optimization map: contours of $|E_{J2}/E_{J1}| = \tau_e/16$ in the $(\sigma, d_0)$ plane at $\theta=1.15$~eV, evaluated at fixed $E_J/h = 8$~GHz. Markers show the fitted Region-I $(d_\mathrm{I},\sigma_\mathrm{I})$ parameters of each device family in this work (Square Set A and B, HAR Sets A, B, and C). The right bar places each family on the resulting $|E_{J2}/E_{J1}|$ scale.
    }
    \label{fig:Fig5}
    \end{figure}

The fitted parameters in Table~\ref{tab:Table1} can then be used to generate the corresponding transparency distribution for a typical junction of a desired area. As an example, we choose the Square Set A 30~Torr$\cdot$min parameters (Table~\ref{tab:Table1}) to calculate representative histograms of $\rho_\mathrm{I}(\tau)$ (Fig.~\ref{fig:Fig5}a). These histograms are used to estimate $I_1$ and $I_2$ for square junctions across the range $0.01 \le A_\mathrm{jj} \le 2.0~\mu\mathrm{m}^2$. In large junctions, although the total number of channels in Region II far exceeds that in Region I, all 50 of the most transparent channels still lie in Region I. Figure~\ref{fig:Fig5}b quantifies the fractional contribution of the top 10, 50, and 1000 channels to $I_1$ and $I_2$. Surprisingly, the top 10 contribute $\gtrsim90\%$ of $I_2$ at every area. Even in the largest devices, where Region II supplies $\sim$86\% of the channels and nearly all of $I_1$, $I_2$ remains carried by a small number of Region-I channels. This dominance of $I_2$ by so few channels is consistent with the small, slowly varying $N_e$. For a more direct comparison with transmon results, we plot the experimental $\tau_e/16=E_{J2}/E_{J1}$ data at $E_J/h = 8$~GHz and overlay contours of the simulated $E_{J2}/E_{J1}$ on the $(d_0,\sigma)$ plane in Fig.~\ref{fig:Fig5}c. Each fitted curve reduces to a single point in this plane, with all $|E_{J2}/E_{J1}|$ values falling between $10^{-4}$ and $2\times10^{-3}$. These values are of similar magnitude to those in earlier work, but point to an alternative interpretation~\cite{kim2025emergentharmonicsjosephsontunnel,Willsch2024}. Kim \textit{et al.}\ carefully attributed the harmonics primarily to parasitic series inductance, under the assumption of a fixed, area-independent junction contribution ($\tau_e = \mathrm{constant}$). Our methodology shows that the HAR geometries used in that study represent a special case in which the linear trend in $|E_{J2}/E_{J1}|$ can equally arise from non-ergodic scaling within the junction itself.

The area dependence observed provides a design handle for superconducting circuits, showing that higher Josephson harmonic content is inherently set by junction geometry for fixed oxide properties. The ensemble $\rho(\tau)$ is a property of the oxide, and as more junctions are placed in a circuit the law of large numbers dictates that the extreme $\tau$ values will be sampled. In systems with thousands of qubits, designers will need to account for those atypical junctions with very transparent channels sampled from the extreme right tail of $\rho(\tau)$ and with anomalously large higher harmonic content. The only way to eliminate these channels is to reduce oxide thickness variability so that $\rho(\tau)$ never develops a fat tail and remains self-averaging.

The data indicate that this disorder is inherited from the metal rather than generated by the oxidation, since $\sigma_\mathrm{I}$ is independent of oxidation exposure while $d_\mathrm{I}$ grows with it (Table~\ref{tab:Table1}). Likewise, $l_\mathrm{th}$ is exposure-independent yet differs between processes. The fitted roughness $\sigma_\mathrm{I}\approx0.22$--$0.29$~nm is comparable to a single atomic step of Al (0.20--0.23~nm), whereas $\sigma_\mathrm{II}\lesssim0.14$~nm is sub-monolayer, suggesting that Region I is oxide grown over a stepped or faceted bottom-electrode surface and Region II is oxide grown over flatter terraces. Grain boundaries are the natural candidate for such a surface, and cross-sectional TEM and atomistic simulations indicate that oxide grown over boundaries between crystallographic domains and along film edges differs in thickness and structure from that over the domain interiors~\cite{oxide_growth_Acta,cyster2021simulating,Kennedy2025,wolff2026structuralcontroltwoleveldefect,zhu2026devicevariabilityjosephsonjunctions,Zeng_2015}. Sidewall roughness of the climbing edge is a second candidate, but the rising-edge overlap has no appreciable effect on $\tau_e$ (HAR~C). Moreover, a purely perimeter or edge contribution would grow as $l+w$ and make $I_2$ increase at least as $A_\mathrm{jj}^{1/2}$, whereas we observe saturation. Rather than an edge effect, the fixed size of $A_\mathrm{I}$ may therefore reflect a systematic roughness of the Al surface that arises when electrodes are fabricated with critical dimensions below $l_\mathrm{th}$.

Finally, because the measurement and analysis techniques presented are insensitive to series inductance, our area-dependent subgap characterization provides a geometry- and process-agnostic screen for the harmonic content intrinsic to Josephson junctions, and a powerful methodology for comparing junction fabrication processes. Applied consistently, the optimization map of Fig.~\ref{fig:Fig5}c enables such comparisons across processes, provided the geometric trends identified in this work are taken into account.

\backmatter

\bmhead{Supplementary information}

Supplementary Information is available for this paper.

\bmhead{Acknowledgements}

We thank John Teufel, Matt Matheny and Bastien Dassonneville for helpful discussions; Simon Luong, Lawrence Wong and Matthew Fling for hardware and cryogenics support; and the staff across the AWS Center for Quantum Computing (CQC) who enabled this work.

\section*{Declarations}

\begin{itemize}
\item \textbf{Funding:} Not applicable.
\item \textbf{Competing interests:} The authors declare no competing interests.
\item \textbf{Data availability:} Data supporting the findings of this study are available from the corresponding author upon reasonable request.
\item \textbf{Author contribution:} D.J.P. contributed to the experimental design, developed the instrumentation and model, performed the primary measurements, analyzed the data, and wrote the manuscript. A.B., M.C., and G.R. contributed to the analysis, played a critical role in contextualizing the results and framing the discussion, with A.B. and M.C. also helping to draft the manuscript. M.P., H.L., and C.C. contributed to the experimental design and measurements, and provided editorial support.  Y.S. provided supervision and contributed to the experimental design and manuscript editing.

\end{itemize}

\pdfbookmark[0]{References}{references}
\bibliography{citations}

\clearpage
\pdfbookmark[0]{Supplementary Information}{supplementary}
\addtocontents{toc}{\protect\setcounter{tocdepth}{3}}
\setcounter{section}{0}\setcounter{figure}{0}\setcounter{table}{0}\setcounter{equation}{0}
\renewcommand{\thefigure}{S\arabic{figure}}
\renewcommand{\thetable}{S\arabic{table}}
\renewcommand{\theequation}{S\arabic{equation}}
\renewcommand{\thesection}{S\arabic{section}}
\makeatletter
\def\theHfigure{S\arabic{figure}}
\def\theHtable{S\arabic{table}}
\def\theHequation{S\arabic{equation}}
\def\theHsection{S\arabic{section}}
\makeatother
\begin{center}
{\Large\bfseries Supplementary Information}\\[6pt]
{\large Non-ergodic Area Scaling of Andreev Transport in Al/AlO$_x$/Al Josephson Junctions}
\end{center}

This document provides supplementary information for ``Non-ergodic Area Scaling of Andreev Transport in Al/AlO$_x$/Al Josephson Junctions,'' including fabrication details, measurement setup, supplemental DC transport data, details of the direct tunneling model with oxide thickness disorder, and model fitting procedures.

\tableofcontents

\pdfbookmark[1]{Symbol Definitions}{symbols}
\section*{Symbol Definitions}

\begin{itemize}
    \item $\boldsymbol{d}$: oxide barrier thickness (nm).
    \item $\boldsymbol{d_0}$: mean oxide thickness (nm) of a generic thickness distribution $\rho(d)$, used where no region is specified; the fitted Region-I and Region-II values are $d_\mathrm{I}$ and $d_\mathrm{II}$.
    \item $\boldsymbol{\sigma}$: standard deviation of the oxide thickness distribution (nm).
    \item $\boldsymbol{\theta}$: average barrier height (eV) of the AlO$_x$ insulator.
    \item $\boldsymbol{\kappa = \sqrt{2m_e\theta}/\hbar}$: tunneling wavefunction decay constant (nm$^{-1}$); the transmission probability decays as $e^{-2\kappa d}$.
    \item $\boldsymbol{\omega = \kappa\sigma}$: dimensionless tunneling disorder parameter. $\omega = 0.5$ is the fat-tail onset (Eq.~\ref{eq:S11}); for $\omega \gtrsim 0.8$ the sampled $\tau_e$ becomes size-dependent over the experimental range, and for $\omega > 1$ the thinnest channels exponentially dominate the higher moments (extreme-value regime).
    \item $\boldsymbol{\rho(d)}$: oxide thickness probability distribution.
    \item $\boldsymbol{n_\mathrm{ch}}$: areal density of independent tunneling channels ($\mu$m$^{-2}$).
    \item $\boldsymbol{N_\mathrm{ch} = A_\mathrm{jj} \cdot n_\mathrm{ch}}$: total number of channels in a junction.
    \item $\boldsymbol{M_k = (1/N_\mathrm{ch})\sum_i \tau_i^k}$: $k$-th moment of the transparency distribution.
    \item $\boldsymbol{\tau_e = M_2/M_1}$: effective transparency.
    \item $\boldsymbol{N_e = N_\mathrm{ch}\,M_1^2/M_2}$: effective number of channels.
    \item $\boldsymbol{m, s}$: log-normal shape parameters, $m = \ln d_0 - s^2/2$, $s = \sqrt{\ln(1+\sigma^2/d_0^2)}$.
    \item $\boldsymbol{A_\mathrm{jj}}$: total junction area ($\mu$m$^2$).
    \item $\boldsymbol{l_\mathrm{th}}$: threshold length ($\mu$m); junctions with $\min(l,w)<l_\mathrm{th}$ behave as pure Region I.
    \item $\boldsymbol{t_\mathrm{b}}$: bottom-electrode thickness.
    \item $\boldsymbol{A_\mathrm{th} = l_\mathrm{th}^2 + 2\,l_\mathrm{th}\,t_\mathrm{b}}$: area, sidewalls included, of a square junction of side $l_\mathrm{th}$; the Region-I area saturates at $A_\mathrm{th}$.
    \item $\boldsymbol{A_\mathrm{I},\; A_\mathrm{II}}$: areas of Regions I and II (main text, Eq.~\ref{eq:4}); $A_\mathrm{I}=A_\mathrm{jj}$ for $\min(l,w)<l_\mathrm{th}$ and $A_\mathrm{I}=A_\mathrm{th}$ otherwise, $A_\mathrm{II}=A_\mathrm{jj}-A_\mathrm{I}$.
    \item $\boldsymbol{d_\mathrm{I},\; \sigma_\mathrm{I}}$: mean thickness and disorder of Region I (the disordered, bounded region).
    \item $\boldsymbol{d_\mathrm{II},\; \sigma_\mathrm{II}}$: mean thickness and disorder of Region II (the more uniform region).
    \item $\boldsymbol{\theta_\mathrm{I},\; \theta_\mathrm{II}}$: barrier heights of Regions I and II (eV).
    \item $\boldsymbol{\phi}$: superconducting phase difference across the junction.
    \item $\boldsymbol{a_1,\; a_2}$: amplitudes of the first and second Fourier harmonics of the current-phase relation (CPR).
    \item $\boldsymbol{|a_2/a_1|}$: CPR skew ratio; $|a_2/a_1| \approx \tau_e/8$ to leading order.
\end{itemize}

\section{Fabrication}
\label{sec:S1}

See Table~\ref{tab:TableS1}. All junctions are Manhattan-style Al/AlO$_x$/Al junctions, shadow-evaporated in a UHV electron-beam
evaporator (Angstrom Engineering, HAR~A in a Plassys) on high-resistivity Si. The tunnel barrier was formed by static oxidation in O$_2$; the oxidation condition is quoted as the
exposure $\mathcal{E}$, the product of the static O$_2$ pressure and the oxidation time (Torr$\cdot$min), which sets the barrier thickness and specifies the process. For $R_\mathrm{n} A_\mathrm{jj}$ the median value is given and $n$ the number of junctions analysed. Individual $R_\mathrm{n} A_\mathrm{jj}$ values span $\approx$320--4190~$\Omega\,\mu$m$^2$, which covers standard fabrication
conditions reported in the literature~\cite{Arute2019,Gambetta2017,Barends2013,place2021new,Krantz2019}.
Areas are $A_\mathrm{jj} = l\,w + 2\,l\,t_\mathrm{b}$, with $w$ the bottom- and $l$ the
top-electrode width and $t_\mathrm{b}$ the bottom-electrode thickness; the second term is
the overlap of the top electrode with the two sidewalls of the bottom electrode, over which the top electrode climbs. $l, w$ are design values except for Square Set~A, whose face dimensions were measured by
SEM on a duplicate wafer, since the as-fabricated dimensions differed from design. The tabulated
$A_\mathrm{jj}$ and $R_\mathrm{n} A_\mathrm{jj}$ include the sidewall term.

\begin{table}[!tbp]
\centering
\caption{Fabrication and geometry of the measured junction sets.}
\label{tab:TableS1}
\footnotesize
\setlength{\tabcolsep}{3pt}
\begin{tabular}{llccrc}
\toprule
Set & Junction geometry ($\mu$m) & $\mathcal{E}$ (Torr$\cdot$min) & $A_\mathrm{jj}$ ($\mu$m$^2$) & $R_\mathrm{n} A_\mathrm{jj}$ ($\Omega\,\mu$m$^2$) & $n$ \\
\midrule
Square A & $l = w = 0.15$--$1.2$ & 15 & 0.054--1.71 & 676 & 8 \\
Square A & $l = w = 0.15$--$1.2$ & 30 & 0.054--1.71 & 918 & 16 \\
Square A & $l = w = 0.15$--$1.2$ & 120 & 0.054--1.71 & 1906 & 17 \\
Square A & $l = w = 0.15$--$1.2$ & 240 & 0.054--1.71 & 3090 & 16 \\
\midrule
Square B & $l = w = 0.1$--$1.0$ & 4 & 0.018--1.08 & 340 & 7 \\
Square B & $l = w = 0.2$--$1.2$ & 87 & 0.055--1.53 & 1497 & 10 \\
Square B & $l = w = 0.15$--$1.2$ & 480 & 0.034--1.53 & 3474 & 12 \\
\midrule
HAR A & $l = 0.2$ or 0.3, $w = 0.15$--$2.0$ & 100 & 0.045--0.62 & 1495 & 22 \\
HAR B & $l = 0.2$ or 0.3, $w = 0.15$--$2.0$ & 100 & 0.045--0.62 & 1333 & 16 \\
HAR C & $l = 0.1$--$1.2$, $w = 0.2$--$0.75$ & 96 & 0.038--0.45 & 1243 & 10 \\
\bottomrule
\end{tabular}
\end{table}

\begin{figure}[!htb]
    \centering
    \includegraphics[width=1\linewidth]{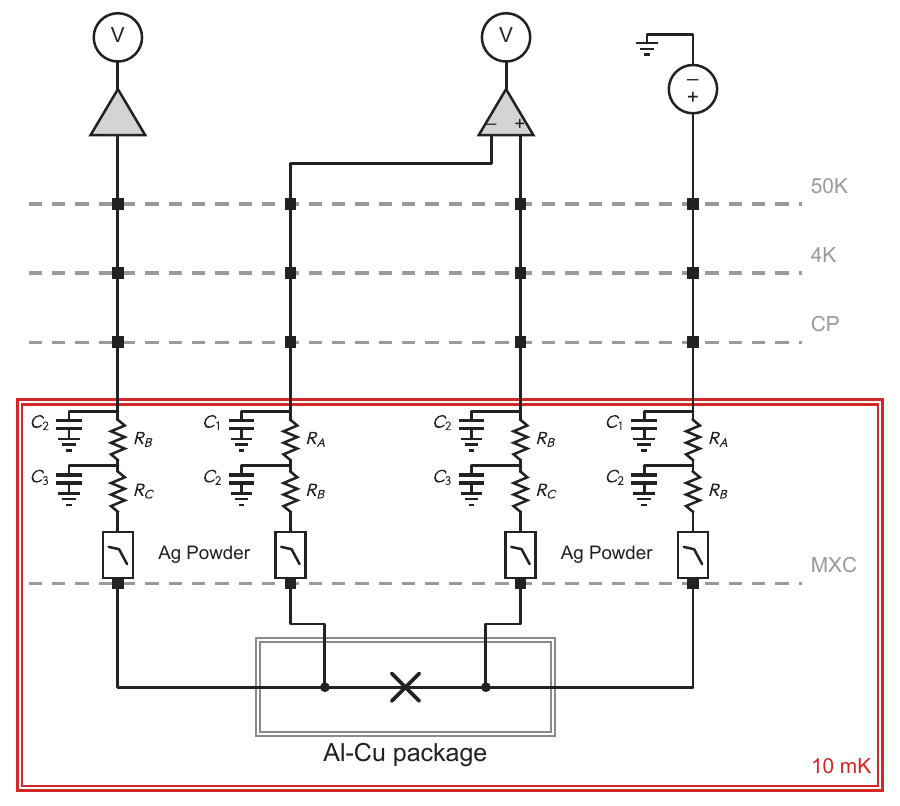}
    \caption{Schematic of the cryogenic DC measurement circuit. Temperature
    stages are indicated by horizontal dashed lines (50~K, 4~K, cold
    plate~(CP), and mixing chamber~(MXC) at $\sim$10~mK, enclosed by
    the red box). The junction (cross) is measured in a four-probe
    configuration: the DC bias line (right) and the current-measurement
    line (left) drive the two ends of the junction wire, and the two
    inner lines sense the voltage across it. Each line passes through a
    two-stage $RC$ low-pass filter at MXC with component values
    $R_A = 50~\Omega$, $R_B = 100~\Omega$, $R_C = 220~\Omega$,
    $C_1 = 47$~nF, $C_2 = 33$~nF, $C_3 = 15$~nF, followed by a Basel
    MFT silver-powder filter. Room-temperature electronics are a Basel
    LNHR DAC (DC voltage source), an SRS~SR560 voltage preamplifier and
    an SRS~SR570 current preamplifier, each read out by a Keithley~7510
    digital multimeter (V). All DC wiring uses shielded twisted-pair
    cable. The sample package (gray) has a copper base and an aluminum
    light-tight lid.}
    \label{fig:FigS1}
\end{figure}
\section{Measurement Setup and Wiring}

All measurements were performed in a dilution refrigerator (Bluefors
LD) at a base temperature of approximately 10~mK. Junctions were
mounted in an Al/Cu light-tight sample package. A schematic of the DC measurement circuit is
shown in Fig.~\ref{fig:FigS1}.

DC bias lines were heavily filtered using a combination of Basel MFT
silver-powder filters and modified QDevil $RC$ filters, configured in
two stages. The low-pass $RC$ component values were chosen such that
the total line resistance satisfies
$R_\mathrm{line} < R_\mathrm{jj}$ while maintaining a 3~dB cutoff
frequency below 50~kHz.

A Basel LNHR DAC served as the low-noise DC voltage source. Voltage was measured using an SRS~SR560 low-noise voltage
preamplifier followed by a Keithley~7510 digital multimeter. Current
was measured using an SRS~SR570 low-noise current preamplifier and a
second Keithley~7510.

\section{Supplemental DC Transport Data}
\label{sec:S3}

\begin{figure}[!tbp]
    \centering
    \includegraphics[width=1\linewidth]{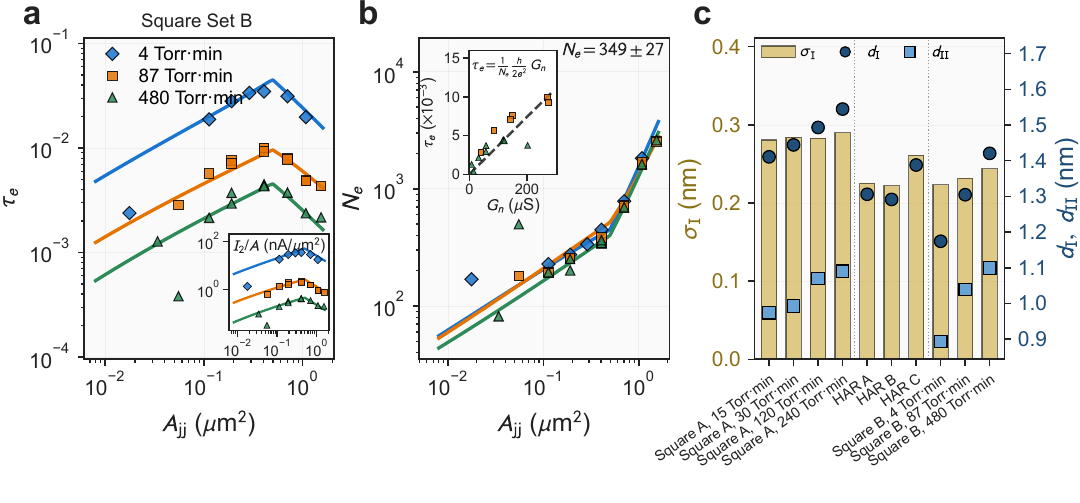}
    \caption{Supplemental fit example from an independent fabrication batch (Square Set B), demonstrating that the non-monotonic $\tau_e$--$A_\mathrm{jj}$ trend reproduces across processes. Fits (solid lines) use $\theta_\mathrm{I}=\theta_\mathrm{II}=1.15$~eV.
    \textbf{a},~$\tau_e$ versus $A_\mathrm{jj}$ for oxidation exposures of 4, 87, and 480~Torr$\cdot$min. The single smallest-area 4~Torr$\cdot$min die ($A_\mathrm{jj}=0.018~\mu\mathrm{m}^2$, drawn) is excluded from the joint fit as an outlier. Inset: $I_2/A_\mathrm{jj}$ versus $A_\mathrm{jj}$, showing the saturation of $I_2$ above $A_\mathrm{jj}=A_\mathrm{th}$.
    \textbf{b},~$N_e$ versus $A_\mathrm{jj}$. Inset: $\tau_e$ versus $G_n$ for the small-junction regime (87 and 480~Torr$\cdot$min dies), whose inverse slope $\tau_e=\frac{h}{N_e 2e^2}G_n$ gives a constant $N_e = 349\pm27$.
    \textbf{c},~Extracted Region-I disorder $\sigma_\mathrm{I}$ (bars) and mean Region-I and Region-II oxide thicknesses $d_\mathrm{I},d_\mathrm{II}$ (markers) for the Square Set A exposure series, HAR Sets A, B, and C, and Square Set B. The HAR Sets are fit as pure Region I, so no $d_\mathrm{II}$ is shown for them.}
    \label{fig:FigS2}
\end{figure}

Every type of junction analyzed showed the geometry- and $A_\mathrm{jj}$-dependent behavior. Here we show further examples, with fits, which demonstrate that the phenomenology remains the same, but the absolute magnitudes of $\tau_e,N_e$ can depend on the individual fabrication process. Fig.~\ref{fig:FigS2} shows data from Square Set B with oxidation exposures of 4, 87, and 480~Torr$\cdot$min.

\subsection{\texorpdfstring{$\tau_e$ and $N_e$}{tau-e and N-e} versus oxidation exposure}

\begin{figure}[!tbp]
    \centering
    \includegraphics[width=1\linewidth]{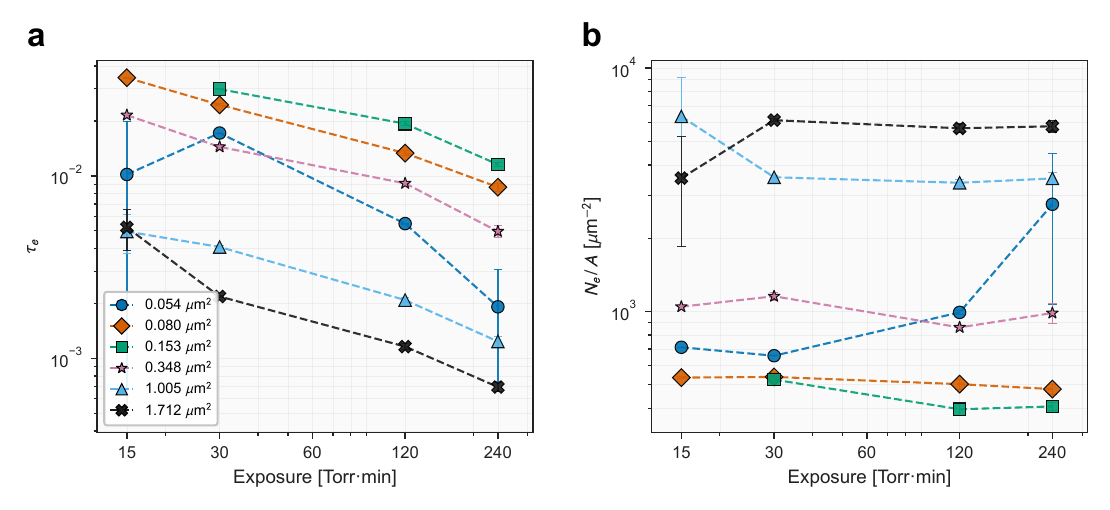}
    \caption{Effective transparency ($\tau_e=I_2/I_1$) and effective
    channel number ($N_e$) versus oxidation exposure.
    \textbf{a},~$\tau_e$ versus oxidation exposure $\mathcal{E}$ on log--log axes
    for Square Set A junction areas from
    $0.054~\mu\mathrm{m}^2$ to $1.71~\mu\mathrm{m}^2$.
    \textbf{b},~Channel density $N_e/A_\mathrm{jj}$ versus oxidation
    exposure on log--log axes. No consistent trend is observed. Variation
    for the smallest area ($A_\mathrm{jj}=0.054~\mu\mathrm{m}^2$) is
    attributed to large propagated error in $I_2$ due to its small
    amplitude.}
    \label{fig:FigS3}
\end{figure}

\begin{figure}[!tbp]
    \centering
    \includegraphics[width=0.5\linewidth]{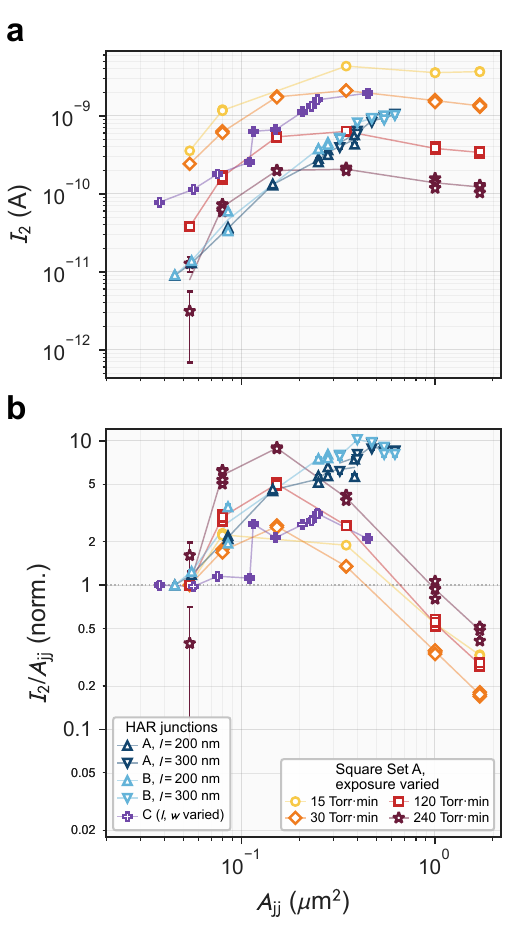}
    \caption{Subgap current $I_2$ versus junction area for the rectangular and square junction families.
    \textbf{a},~Raw $I_2$ versus $A_\mathrm{jj}$, one marker per junction; thin lines connect per-area medians.
    \textbf{b},~Subgap current density $I_2/A_\mathrm{jj}$, with each series normalized to its value at the smallest measured area (within HAR Sets A and B, the $l=300$~nm sweep shares its own set's $l=200$~nm anchor, so each set reads as one band); the dotted line marks the reference level of 1. Compared with the near-linear $I_1\propto A_\mathrm{jj}$ (see $G_n$ vs $A_\mathrm{jj}$ in Fig.~\ref{fig:Fig3}a,d), $I_2/A_\mathrm{jj}$ varies by more than an order of magnitude and dominantly contributes to $\tau_e(A_\mathrm{jj})$.
    }
    \label{fig:FigS4}
\end{figure}

Figure~\ref{fig:FigS3}a plots $\tau_e=I_2/I_1$ (see
Fig.~\ref{fig:Fig2}b of the main text for definitions of $I_{1,2}$) versus
oxidation exposure $\mathcal{E}$ for fixed junction area. For all areas, the observed
power-law dependence is consistent with a Mott--Cabrera oxidation
process~\cite{PhysRevLett.112.146103,Cabrera1949}, wherein the oxide
thickness ($d_0$) grows logarithmically with exposure. Substituting
$d_0 = a + b\ln \mathcal{E}$ into an exponential tunneling model,
$\tau_e \propto e^{-2\kappa d_0}$, yields the power law
$\tau_e \propto \mathcal{E}^{-2\kappa b}$,
which linearizes on the log--log axes of Fig.~\ref{fig:FigS3}a.
Here $b$ is the logarithmic growth coefficient of the oxide, with units
of length, so that the exponent $2\kappa b$ is dimensionless. Fitting the
Region-I thicknesses $d_\mathrm{I}$ of main-text Table~\ref{tab:Table1} against $\ln \mathcal{E}$
gives $b = 0.046$~nm, i.e., 0.11~nm of additional oxide per decade of exposure
(Region II gives 0.045~nm, or 0.10~nm per decade, essentially the same), hence a predicted exponent $2\kappa b = 0.50$. This agrees with the
measured slopes $d\ln\tau_e/d\ln \mathcal{E}$, which have a median of
$-0.50$ across the six areas.
This result serves as confirmation of fabrication consistency between
sample lots, which is critical for reliably extracting the conclusions
presented in the main text.

Figure~\ref{fig:FigS3}b plots $N_e/A_\mathrm{jj}$ for fixed junction areas as a
function of oxidation exposure. Many of the 120 and 240~Torr$\cdot$min data points for
the smallest devices have extremely small $I_2$ values, which propagate
into correspondingly large error bars in $N_e$. No correlation is
observed between $N_e$ and oxidation exposure over the range of areas and
exposures covered in this study.

\section{Determining the form of \texorpdfstring{$\rho(d)$}{rho(d)} based on the area dependence of \texorpdfstring{$\tau_e, N_e$}{tau-e, N-e}}
\label{sec:S4}

The oxide thickness distribution in a Josephson junction has been evaluated with TEM metrology and described by both normal~\cite{GaussianOxide,cyster2021simulating} and log-normal distributions~\cite{zhu2026devicevariabilityjosephsonjunctions}, with the latter argued to be more physical due to the presence of thick regions serving as an upper tail. The right-skew of the oxide thickness distribution $\rho(d)$ is less critical than the shape of the left tail due to the exponential tunneling dominance of thin channels. TEM on cross-sectional cuts is useful for correlation, but cannot easily distinguish the subtle difference in the extreme limits of the left tail structure. Therefore, we develop an analysis procedure that includes (i) introduction of moment generating functions, (ii) order statistics, (iii) tail enhancement, and (iv) moment convergence using order statistics.

The experimental results in the main text indicate both geometric and area dependencies which are unexpected in ergodic systems. For the square devices, $\tau_e$ is both area-dependent and non-monotonic, while $N_e$ has two distinct regions with different $A_\mathrm{jj}$ dependence. For both small-area junctions and high-aspect-ratio junctions, $\tau_e$ increases with area, while $N_e$ is constant or increases sub-linearly. Here, the experimental results are compared with possible solution spaces of log-normal and normal oxide thickness distributions to test whether each can recreate the $A_\mathrm{jj}$ dependence observed for $\tau_e,N_e$. The overarching goal of this section is to define the correct distribution, understand how this distribution results in the area dependence of the observed $N_e,\tau_e$, and discuss junction quality metrics based on the properties of $\rho(\tau(d))$.

\subsection{Moment-Generating Functions}

Here we derive expressions for the moment-generating function for both normal and log-normal distributions. These functions are useful for evaluating the role of rare tunneling events, which is critical for analyzing subgap $i_\mathrm{dc}$--$v_\mathrm{dc}$ data.

The main text defines the analysis parameters $\tau_e, N_e$ in terms of the experimentally measured $I_1,I_2$. These can also be expanded in terms of the moments of $\rho(\tau)$, $M_k$, as follows:

\begin{equation}
M_k = \frac{1}{N_\mathrm{ch}}\sum_{i=1}^{N_\mathrm{ch}}{\tau_i^k} =\langle\tau^k\rangle  = \int_{0}^\infty{\tau(x)^k\rho(x)dx}.
\end{equation}

$\tau_e, N_e$ can then be written in terms of $M_1,M_2$:

\begin{equation}
N_\mathrm{ch}\,M_1 = N_\mathrm{ch}\langle \tau \rangle = \sum_{i=1}^{N_\mathrm{ch}}{\tau_i} =G_n\frac{h}{2e^2}
\end{equation}
\begin{equation}
N_\mathrm{ch}\,M_2 = N_\mathrm{ch}\langle \tau^2 \rangle =\sum_{i=1}^{N_\mathrm{ch}}{\tau_i^2} =G_{sg}\frac{h}{2e^2}
\end{equation}

\begin{equation}
\tau_e = \frac{M_2}{M_1}
\end{equation}

\begin{equation}
N_e = N_\mathrm{ch}\,\frac{M_1^2}{M_2}
\end{equation}

\subsubsection{Normal distribution}

The normal distribution has the form:
\begin{equation}
\rho_{N}(d)=\frac{1}{\sigma\sqrt{2\pi}}e^{-\frac{(d-d_0)^2}{2\sigma^2}}
\label{eq:S6}
\end{equation}

from which we can calculate the $n^\text{th}$ moment (using $\tau(d)^n = (C/d)^n e^{-2n\kappa d}$, where $C$ is the prefactor of the Simmons transparency, Eq.~\ref{eq:5} of the main text):

\begin{equation}
M_n=C^n\int_{-\infty}^\infty{d^{-n}e^{-\frac{1}{2\sigma^2}[d^2-2(d_0-2n\kappa \sigma^2)d+d_0^2]}}\frac{dd}{\sigma\sqrt{2\pi}}
\label{eq:S7}
\end{equation}

Defining a shifted mean $\mu_n^*=d_0-2n\kappa\sigma^2$, $M_n$ simplifies to:

\begin{equation}
M_n=C^ne^{-2n\kappa d_0+2n^2\kappa^2\sigma^2}\langle d^{-n}\rangle_{\mu_{n}^*}
\label{eq:S8}
\end{equation}

The experimental data are limited to the first two moments of $\rho(\tau)$:

\begin{equation}
M_1\approx \frac{C}{(\mu_1^*)}e^{-2\kappa d_0+2\kappa^2\sigma^2},~~ \mu_1^*=d_0-2\kappa\sigma^2
\label{eq:S9}
\end{equation}

\begin{equation}
M_2\approx \frac{C^2}{(\mu_2^*)^2}e^{-4\kappa d_0+8\kappa^2\sigma^2},~~ \mu_2^*=d_0-4\kappa\sigma^2
\label{eq:S10}
\end{equation}

\subsubsection{Defining an \texorpdfstring{$\omega = 0.5$}{omega = 0.5} criterion, and the validity of
Eqs.~\ref{eq:S9}--\ref{eq:S10}.}
In the equations above, $\tau(d)^n$ weights the thickness distribution by
$e^{-2n\kappa d}$, and the channels that carry $M_n$ are not those near $d_0$ but
those near $\mu_n^*$ which are displaced into the thin tail by
\begin{equation}
\frac{d_0-\mu_n^*}{\sigma}=2n\omega,
\qquad
\frac{\mu_n^*-\mu_{n+1}^*}{\sigma}=2\omega,
\label{eq:S11}
\end{equation}
where the displacement is expressed in units of the disorder $\sigma$ itself.
Equivalently, the standard deviation of $\ln\tau$ $\approx2\kappa\sigma
=2\omega$. Both expressions are functions of $\omega$ only: they contain no
reference to $d_0$, $\theta$, or $N_\mathrm{ch}$. The condition $1-2\omega=0$,
i.e.~$\omega=0.5$, is therefore the point at which each successive moment is
sourced by a subpopulation displaced a full standard deviation from the mean
relative to the previous one. Below it, all accessible moments are set by
channels within $\sim\sigma$ of $d_0$ and the transparency statistics
self-average; above it, $M_2$ is controlled by a rare thin-channel tail whose
sampling depends on $N_\mathrm{ch}$. This is the defining property of the
fat-tail regime and the reason $\omega$ can be used as a figure of merit for the disorder. The result is self-consistent with the size-dependent estimate in the main text also. For $\theta=1.15$~eV ($\kappa = 5.49$~nm$^{-1}$), $\omega=0.5$ corresponds to $\sigma = 0.5/\kappa = 0.091$~nm. $\omega = 0.5$ is therefore the lower limit for any size dependence, although the effect is gradual. Below it, $\tau_e$ is size-independent at every $N_\mathrm{ch}$. The threshold quoted in the main text, $\kappa\sigma\gtrsim0.8$, lies above this limit because the experimental window $10^4<N_\mathrm{ch}<2\times10^6$ is finite. Between the two, the tail exists but is not sampled strongly enough to show over our range of junction sizes.

\begin{table}[!tbp]
\centering
\caption{Small-junction (pure Region-I) best-fit parameters for Fig.~\ref{fig:FigS8}, with barrier height fixed at $\theta_\mathrm{I} = 1.15$~eV ($\kappa = 5.49$~nm$^{-1}$).}\label{tab:TableS2}
\begin{tabular}{l c c c c}
\toprule
Geometry & $d_\mathrm{I}$ (nm) & $\sigma_\mathrm{I}$ (nm) & $\sigma_\mathrm{I}/d_\mathrm{I}$ & $\omega_\mathrm{I}$ \\
\midrule
Square Set A (120 Torr$\cdot$min) & 1.498 & 0.285 & 0.190 & 1.56 \\
HAR A+B ($l$ = 200~nm, pooled)     & 1.302 & 0.226 & 0.173 & 1.24 \\
\bottomrule
\end{tabular}
\end{table}

\begin{figure}
    \centering
    \includegraphics[width=\textwidth]{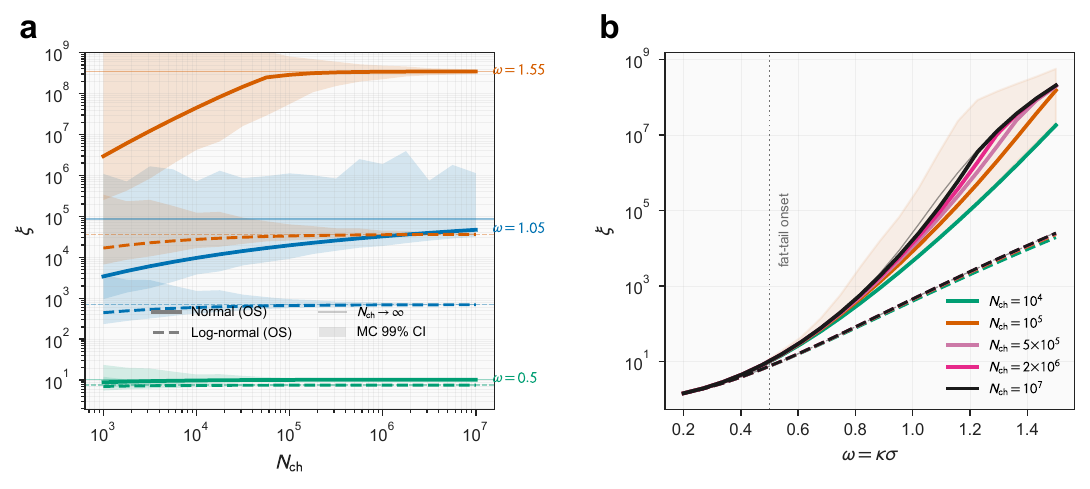}
    \caption{Tail enhancement factor $\xi=M_2/\tau_0^2$ comparing
    order-statistics (OS) summation to the exact $N_\mathrm{ch}\to\infty$
    integral, for normal (solid) and log-normal (dashed) thickness distributions with $d_0 = 1.269$~nm and $\theta = 1.15$~eV, using the numerical floor $d\ge0.05$~nm and the unitarity cap $\tau\le1$ of all calculations in this work. Shaded bands denote the 99\% confidence interval (0.5--99.5 percentile) of the Monte Carlo simulations. Without spatial correlations, the variation around $\mathbb{E}\bigl[d_{(k)}\bigr]$ results in overestimated tail amplification for small $N_\mathrm{ch}$.
    \textbf{a},~Convergence with channel count $N_\mathrm{ch}$ at fixed
    $\omega = 0.5$, $1.05$, and $1.55$; thin horizontal lines mark the
    $N_\mathrm{ch}\to\infty$ limit. At the onset value $\omega = 0.5$ the
    curves are flat, i.e.\ $\xi$ is already converged at $N_\mathrm{ch}=10^3$. At $\omega = 1.55$ the normal-distribution curve saturates early because its thinnest channels reach the cap $\tau=1$; the capped sub-population then carries $M_2$.
    \textbf{b},~Dependence on disorder strength $\omega$ at fixed $N_\mathrm{ch}$,
    with $1000$ Monte Carlo realizations drawn from the corresponding distribution.
    The dotted vertical line marks the fat-tail onset $\omega = 0.5$.
    The dimensionless parameter $\omega = \kappa\sigma$ is the natural
    scale for the normal distribution; log-normal curves are plotted at
    matched variance (i.e.\ the same physical $\sigma$), with shape
    parameter $s = \sqrt{\ln(1 + \sigma^2/d_0^2)}$.  At large $\omega$
    the two distributions diverge because the log-normal tail vanishes
    as $d\to 0$, suppressing the thin-barrier enhancement that dominates
    $M_2$.}
    \label{fig:FigS5}
  \end{figure}

\begin{figure}[!tbp]
    \centering
    \includegraphics[width=1\linewidth]{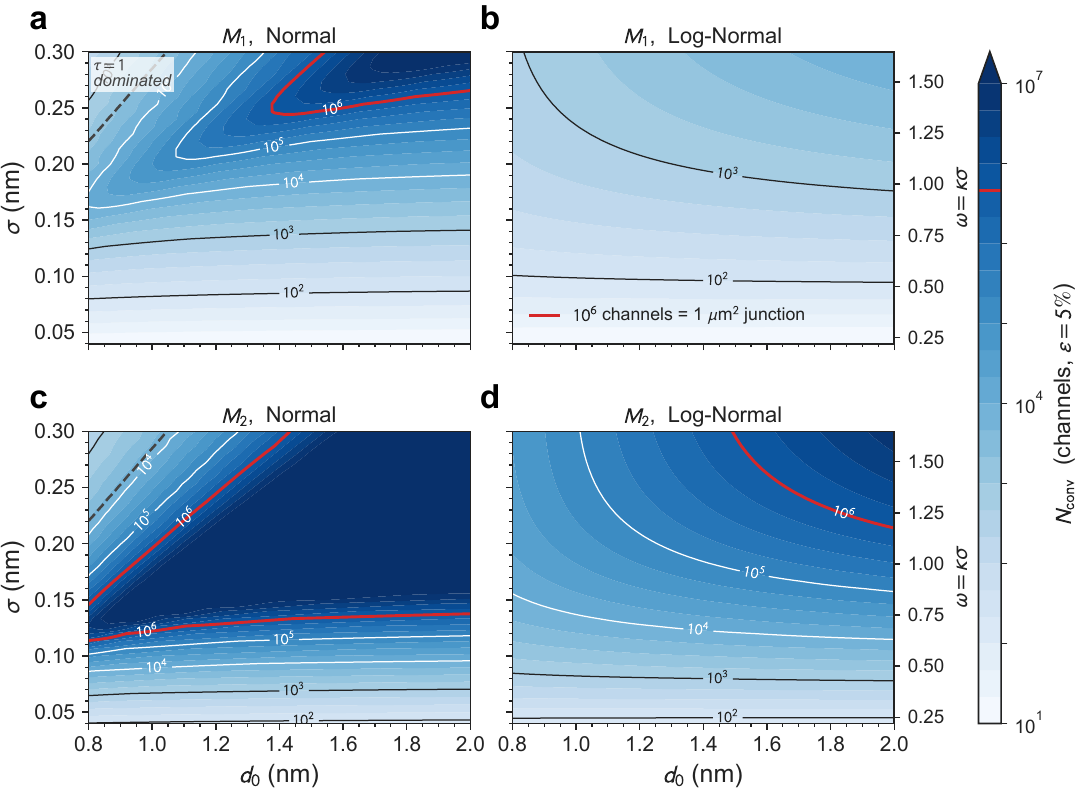}
    \caption{Required number of channels $N_\mathrm{ch}=N_\text{char}^{(n)}/\epsilon^2$ for the partial moments $M_1^p$ \textbf{a},\textbf{b} and $M_2^p$ \textbf{c},\textbf{d} to converge to within a fractional tolerance $\epsilon=0.05$ of their asymptotic values, shown as a function of $(\sigma,d_0)$ for normal \textbf{a},\textbf{c} and log-normal \textbf{b},\textbf{d} thickness distributions. The red contour marks $N_\mathrm{ch}=10^6$, i.e.\ a $1~\mu\text{m}^2$ junction at $n_\mathrm{ch}=10^6~\mu\text{m}^{-2}$. $M_1$ converges according to its variance, $N_\text{char}^{(1)}=M_2/M_1^2-1$; $M_2$ convergence is set by $N_\text{char}^{(2)}=M_{4}/M_2^2-1$. For the log-normal distribution, typical moments are converged for all relevant $A_\mathrm{jj}$; $M_2$ is converged for devices with $A_\mathrm{jj}\,n_\mathrm{ch}\gtrsim10^4$--$10^5$. For the normal distribution, $M_1$ depends on $\sigma$ and $d_0$, with convergence occurring across a large range $A_\mathrm{jj}\,n_\mathrm{ch}\approx10^5$--$10^7$. For disordered oxides ($\sigma_\mathrm{I}>0.25$~nm), small areas may therefore be expected to have $G_n/A_\mathrm{jj}$ depend on $A_\mathrm{jj}$. $M_2$ effectively never converges for these devices, consistent with experimental results where $N_e,\tau_e$ vary across the entire range of junction areas. The dashed line in the normal-distribution panels bounds the corner of parameter space in which more than 0.1\% of all channels sit at the unitarity cap $\tau=1$; there the capped sub-population dominates the moments, and the rapid convergence reflects the cap rather than the sampling of the normal tail. All fitted devices lie far outside this corner.
    }
    \label{fig:FigS6}
\end{figure}

The same logic sets the limits of Eqs.~\ref{eq:S9} and
\ref{eq:S10}. They are second-order Laplace approximations about
$\mu_n^*$. Checked against direct
numerical integration of Eq.~\ref{eq:S7} at $d_0=1.269$~nm, the
closed form for $M_2$ is accurate to 2\% at $\omega=0.5$ and 6\% at
$\omega=0.7$, but the expression fails entirely by $\omega\approx1.32$, where
$\mu_2^*=d_0-4\kappa\sigma^2$ passes through zero and the $1/(\mu_2^*)^2$
prefactor diverges. At the fitted Region-I disorder $\omega_\mathrm{I}\approx1.5$
(Table~\ref{tab:TableS2}) $\mu_2^*$ is negative, so
Eqs.~\ref{eq:S9}--\ref{eq:S10} cannot be evaluated at the
parameters of these devices and are retained only to expose the $\mu_n^*$
structure.
Every moment, $\xi$, $\tau_e$, and $N_e$ value quoted in this work
is instead computed from either the order-statistics sum
(Eq.~\ref{eq:S18}), from Monte Carlo sampling, or from
direct numerical integration of Eq.~\ref{eq:S7}. All of these
calculations apply the same two cutoffs, a numerical floor $d\ge0.05$~nm on
the thickness and $\tau\le1$ on the transparency. At every
fitted $(d_0,\sigma)$ and every measured $N_\mathrm{ch}$, the smallest order
statistic lies well above one monolayer, so no fitted parameter depends on
either cutoff. The far-field illustrations
(Figs.~\ref{fig:FigS5} and \ref{fig:FigS6}) extrapolate
to channel counts and disorders at which the thin tail reaches $\tau=1$. There the capped and perfectly transmitting sub-population carries the higher
moments, with the region where it dominates marked in
Fig.~\ref{fig:FigS6}.

\subsubsection{Log-normal distribution}

For the log-normal distribution, we have:

\begin{equation}
\rho_{LN}(d)=\frac{1}{d\,s\sqrt{2\pi}}\,\exp\!\left(-\frac{(\ln d-m)^2}{2s^2}\right)
\end{equation}
For the log-normal to have the same mean and variance as the normal distribution, $e^{m+s^2/2}=d_0$ and $(e^{s^2}-1)e^{2m+s^2}=\sigma^2$, giving $m\approx \ln d_0$ and $s\approx \sigma/d_0$ when $s$ is small. The $n^\text{th}$ moment is
\begin{equation}
M_n=\frac{C^n}{s\sqrt{2\pi}}\int_{-\infty}^{\infty}{e^{f_n(u)}du}
\label{eq:S13}
\end{equation}

with

\begin{equation}
f_n(u)=-nu-2n\kappa e^u-\frac{(u-m)^2}{2s^2}
\label{eq:S14}
\end{equation}

\subsection{Order Statistics and Partial Moments}

The expectation value (first moment) of the $i^{th}$ order statistic (OS, or ordered thickness) taken from a set of $N_\mathrm{ch}$ oxide thicknesses with given thickness distribution $\rho(d)$ is given by:

\begin{equation}
    \mathbb{E}[d_{(i)}]=\int_{t_{ox}=0}^{t_{ox}=\infty}t_{ox}\cdot f_{d_{(i)}}(t_{ox})dt_{ox}
\end{equation}

where $t_{ox}=d$ is the oxide thickness, $1\le i\le N_\mathrm{ch}$, and $f_{d_{(i)}}$ is the PDF of the $i^{th}$ order statistic. For a normal distribution this is estimated using the quantile (probit) approximation~\cite{david2003order}:

\begin{equation}
    \mathbb{E}\bigl[d_{(i)}\bigr] \approx d_0+\sigma\,Q\!\left(\frac{i}{N_\mathrm{ch}+1}\right)
\end{equation}

where $Q(\cdot)$ denotes the standard-normal quantile (probit) function.

For a log-normal distribution, $\mathbb{E}\bigl[d_{(i)}\bigr]$ is then given by:

\begin{equation}
\begin{split}
    \mathbb{E}\bigl[d_{(i)}\bigr] &\approx e^{m+s\,Q\left(\tfrac{i}{N_\mathrm{ch}+1}\right)}\\
    s &= \sqrt{\ln (1+\sigma^2/d_0^2)}\\
    m &= \ln d_0 -s^2/2.
  \end{split}
\end{equation}

Mean values of the partial moments can then be calculated for $N_\mathrm{ch}$ ($A_\mathrm{jj}$) using order statistics:
\begin{equation}
    M_n^{p}=\sum_{k=1}^{N_\mathrm{ch}}{\tau^n(\mathbb{E}[d_{(k)}])}
  \label{eq:S18}
\end{equation}

\subsection{Tail Enhancement}

At the heart of the non-self-averaging behavior and experimental trends for $\tau_e,N_e$ is the fat tail of the $\rho(\tau)$ distribution. To quantify the rare-event statistics resulting from the tail, we use the above formalism and define an $n^{th}$ order tail-enhancement factor $\xi$ as the ratio of the $n$-th moment to the value expected if all channels had the mean thickness $d_0$:

\begin{equation}
   \xi = \frac{M_n}{\tau(d_0)^n}= \int_0^\infty \left(\frac{d_0}{x}\right)^n e^{-2n\kappa(x - d_0)} \rho(x)dx
\end{equation}

Since $M_n$ is defined per channel (Eq.~\ref{eq:S18} normalized by $N_\mathrm{ch}$), $\xi$ carries no explicit $N_\mathrm{ch}$ factor; its $N_\mathrm{ch}$ dependence enters only through the sampling of the tail. Figure~\ref{fig:FigS5} shows numerical results from Eqs.~\ref{eq:S7},\ref{eq:S13} compared with results from 1000 Monte Carlo (MC) trials and OS calculations. The amount of oxide thickness disorder is represented by the dimensionless $\omega=\kappa\sigma$, which increases with $\sigma$. On one extreme, $\omega\to0$ represents a perfectly uniformly thick oxide layer. In such a case, nearly all channels have the same $\tau \approx \tau_e$ and $N_e\approx N_\mathrm{ch}$. In the other extreme, $\omega$ grows monotonically as $\sigma$ becomes of order $d_0$. This is further shown in the $\tau_e$ - $N_\mathrm{ch}$ simulation in Fig.~\ref{fig:Fig4}, where $d_0$ is fixed but $\sigma$ is swept through representative values for a junction oxide.

For a normal distribution of $d$, the tunneling transmission $\tau(d) = (C/d)e^{-2\kappa d}$ maps a symmetric oxide thickness distribution into a heavily right-skewed transparency distribution. Both the exponential decay and the $1/d$ prefactor amplify the contribution of thin-barrier channels. Even modest oxide roughness $\sigma$ produces a fat-tailed $\rho(\tau)$ whose higher moments are dominated by rare thin channels. The resulting $\xi$ grows as disorder increases.
Conversely, for the log-normal thickness distribution, the saddle point (Eq.~\ref{eq:S14}) never shifts far into the thin-barrier tail ($d\to0$, i.e.\ $u\to-\infty$), and $\xi$ saturates at a fixed value in junctions with $A_\mathrm{jj}>0.01~\mu\text{m}^2$. It is this tail enhancement or fat-tail of the distribution that is largely responsible for the area dependence observed in $\tau_e,N_e$.

The expectation value of probability distributions with fat tails depends on the number of samples. For small devices, the expectation value is channel-number dependent and the value reflects the typical result, because few extreme outliers are sampled. Increasingly, more extreme values (thin channels) occur which dominate $M_k,k \ge 2$, and push the typical solution toward the average. These rare thin-channel transparencies are large enough to rival the \textit{summed} transparency of all remaining channels. With a large enough device, or large enough number of samples, the typical value of $M_1,M_2$ converges to the fully sampled average. The number of channels required to transition from the typical to the average depends on the relative level of oxide disorder, given by $\omega$, as in Fig.~\ref{fig:FigS5}. The fully sampled average regime is well described by the $A_\mathrm{jj}$-independent values given numerically by Eqs.~\ref{eq:S7},\ref{eq:S13}.

Fig.~\ref{fig:FigS5} exemplifies the second-moment tail enhancement $M_2/\tau_0^2$ and its $A_\mathrm{jj}$-dependence for normal distributions with $\omega>1$. The log-normal distribution, however, has orders-of-magnitude smaller $\xi$ and requires only $10^3$--$10^4$ channels for $M_2$ to reach the fully-sampled average. Normal-distributed oxide thicknesses have $\omega$-dependent convergence of the OS simulations to the fully-sampled numerical result. Fig.~\ref{fig:FigS5} indicates that at $\omega\approx1$, $N_\mathrm{ch}\sim 10^7$ is required for convergence, at least 5 times as large as the biggest samples here. At $\omega=1.55$ the earlier saturation is set by the cap $\tau\le1$ rather than by sampling.

\subsection{Numerical Comparison of log-normal and normal \texorpdfstring{$M_n$}{M-n} convergence using order statistics}

The convergence of $M_1^p,M_2^p$ partial moments to their average determines the area dependence of $N_e,\tau_e$. When partial moments are converged, $\tau_e$ will \textit{not} depend on $A_\mathrm{jj}$. While the saturation of $M_2/\tau_0^2$ matches convergence of $M_1^p,M_2^p$ (see Fig.~\ref{fig:FigS5}), here we quantitatively map the $\sigma,d_0$-dependent convergence using the coefficient of variation $\text{CV}(\tau^n)=\frac{\mathrm{SE}(\hat{M}_n)\sqrt{N_\mathrm{ch}}}{M_n}$. If we require $M_n^p$ to lie within $\epsilon$ of $M_n$ with probability 0.68, we have:

\begin{equation}
    N(n,\epsilon)=\frac{\text{CV}(\tau^n)^2}{\epsilon^2}=\frac{N_\text{char}^{(n)}}{\epsilon^2}
\end{equation}

where $N_\text{char}^{(1)}=M_{2}/M_1^2-1 = N_\mathrm{ch}/N_e - 1$ defines $M_1$ convergence, and $N_\text{char}^{(2)}=M_{4}/M_2^2-1$ defines $M_2$ convergence and hence $N_e$ linearity in $N_\mathrm{ch}$, both of which are signatures of convergence to the thermodynamic regime. Fig.~\ref{fig:FigS6} shows the required $N_\mathrm{ch}(\sigma,d_0)$ for different $\epsilon$. If $\sigma,d_0$ are small, normal and log-normal convergence of $M_1,M_2$ occur in the same order of magnitude. However, as $\sigma,d_0$ increase together, the log-normal distribution results in convergence orders of magnitude faster for $M_1$ ($M_2$). The areas probed experimentally all fall into the simulation range that has converged $M_1,M_2$ for $\rho(d)\sim\mathcal{LN}(d_0,\sigma)$. The order of magnitude of $N_\mathrm{ch}$ is not consistent with experimental results, which show no convergence of $\tau_e$.

\subsubsection{Large Junction \texorpdfstring{$\tau_e,N_e$}{tau-e, N-e}}

For large junctions, $M_1,M_2$ converge to their $N_\mathrm{ch}\to\infty$ average values given by numerical integration of Eq.~\ref{eq:S13} and Eq.~\ref{eq:S7}. From Eq.~\ref{eq:S8}:

\begin{equation}
    N_e = N_\mathrm{ch}\frac{M_1^2}{M_2} = N_\mathrm{ch}\,\frac{(\langle d^{-1}\rangle_{\mu_1^*})^2}{\langle d^{-2}\rangle_{\mu_2^*}}\,e^{-4\kappa^2\sigma^2} =\beta N_\mathrm{ch}
\label{eq:S21}
\end{equation}

where $\beta$ is a constant. Similarly, $\tau_e=M_2/M_1$ converges to a constant value:

\begin{equation}
    \tau_e=\frac{\langle d^{-2}\rangle_{\mu_2^*}}{\langle d^{-1}\rangle_{\mu_1^*}}\,C\,e^{-2\kappa d_0+6\kappa^2 \sigma^2}
\end{equation}

both of which follow from Eq.~\ref{eq:S8} and inherit its validity
range $2n\omega\lesssim1$; as noted above, they are not evaluable at the fitted
$\omega_\mathrm{I}\approx1.5$ and the quoted values come from the exact OS/MC/numerical
calculations.

The model reproduces the experimental $N_e$ area dependence. The
constant-$\beta$ limit of Eq.~\ref{eq:S21} is the asymptote reached only
once $M_2$ has converged, which for these disorder parameters requires
$N_\mathrm{ch}\gtrsim10^7$ (Fig.~\ref{fig:FigS6}); over the measured
range $\beta$ instead grows by roughly an order of magnitude. The experimental larger-area trend is dominated by the transition to less-disordered oxide in larger
devices.

\subsubsection{Small-area \texorpdfstring{$\tau_e,N_e$}{tau-e, N-e}: Why is there an area dependence?}

We now derive the expected area dependence of $\tau_e$ using order statistics. From Eq.~\ref{eq:S18}, the partial moments $M_n^p(N_\mathrm{ch})$ are deterministic functions of $N_\mathrm{ch}$. When the channel count increases $N_\mathrm{ch} \to N_\mathrm{ch}+1$, a new thinnest channel always appears with expected thickness:
\begin{equation}
    \mathbb{E}[d_{(1)}] = d_0 + \sigma\,Q\!\left(\frac{1}{N_\mathrm{ch}+2}\right)
  \label{eq:S23}
\end{equation}
for the normal distribution, and

\begin{equation}
    \mathbb{E}[d_{(1)}] = e^{m+sQ\!\left(\frac{1}{N_\mathrm{ch}+2}\right)}
\end{equation}

for log-normal. Simultaneously, all existing quantiles shift from $k/(N_\mathrm{ch}+1) \to k/(N_\mathrm{ch}+2)$. In the small-junction regime the dominant contribution to $\delta\tau_e$ comes from the new extreme channel; the quantile compression of existing channels contributes at $\mathcal{O}(1/N_\mathrm{ch}^2)$. Defining the transparency of this new extreme channel to be $\tau_1 \equiv \tau(\mathbb{E}[d_{(1)}])$, the updated effective transparency $\tau_e'$ is:
\begin{equation}
    \tau_e' = \frac{M_2^p + \tau_1^2}{M_1^p + \tau_1}
\end{equation}
and the expected change in $\tau_e$ per added channel is:
\begin{equation}
    \mathbb{E}[\delta\tau_e] = \tau_e' - \tau_e = \frac{\tau_1(\tau_1 - \tau_e)}{M_1^p + \tau_1} \approx \frac{\tau_1(\tau_1 - \tau_e)}{M_1^p}.
\end{equation}
We can make the approximation $\tau_1 \ll M_1^p$ since even the most transparent single channel carries negligible $I_1$ compared to the sum. $\tau_e$ increases with area whenever the new thinnest channel is more transparent than the current weighted average, i.e., $\tau_1 > \tau_e$.

For the normal distribution, the expected minimum thickness from Eq.~\ref{eq:S23} gives:
\begin{equation}
    \tau_1 = \tau(\mathbb{E}[d_{(1)}]) = \frac{C}{\mathbb{E}[d_{(1)}]}\,e^{-2\kappa\, \mathbb{E}[d_{(1)}]}
\end{equation}
For large $N_\mathrm{ch}$, $Q(1/(N_\mathrm{ch}+2)) \approx -\sqrt{2\ln N_\mathrm{ch}}$, so $\mathbb{E}[d_{(1)}] \approx d_0 - \sigma\sqrt{2\ln N_\mathrm{ch}}$ and:
\begin{equation}
    \tau_1 \sim e^{-2\kappa d_0 + 2\kappa\sigma\sqrt{2\ln N_\mathrm{ch}}}
\end{equation}
This function grows sub-polynomially but without bound in $N_\mathrm{ch}$ (as $e^{2\kappa\sigma\sqrt{2\ln N_\mathrm{ch}}}$). In contrast, for the log-normal distribution:
\begin{equation}
    \mathbb{E}[d_{(1)}] = e^{m - s\sqrt{2\ln N_\mathrm{ch}}}
\end{equation}
with $m = \ln d_0 - s^2/2$ and $s = \sqrt{\ln(1+\sigma^2/d_0^2)}$, giving:
\begin{equation}
    \tau_1 = \frac{C}{\mathbb{E}[d_{(1)}]}\,e^{-2\kappa\, e^{m-s\sqrt{2\ln N_\mathrm{ch}}}}
\end{equation}

The rate $d\tau_e/dN_\mathrm{ch}$, which is an experimentally measurable quantity, trends to zero. Noting that $\mathbb{E}[d_{(1)}^{\text{LN}}] \propto e^{-\sqrt{\ln N_\mathrm{ch}}}$ converges to zero much more slowly than $\mathbb{E}[d_{(1)}^{N}] = d_0 - \sigma\sqrt{2\ln N_\mathrm{ch}}$, $\tau_1^{\text{LN}}$ saturates at modest $N_\mathrm{ch}$ while $\tau_1^{N}$ continues to grow. This causes the log-normal $\tau_e$ to converge to its true average in all but the smallest of junctions, while the normal-distribution $\tau_e$ continues to increase.

Based on the areal trends predicted by the model, only a normal distribution of oxide thickness matches the experimental $A_\mathrm{jj}$ dependence observed in $N_e,\tau_e$.

\subsection{Validating the existence of a small-junction solution space}

\begin{figure}[!tbp]
    \centering
    \includegraphics[width=1\linewidth]{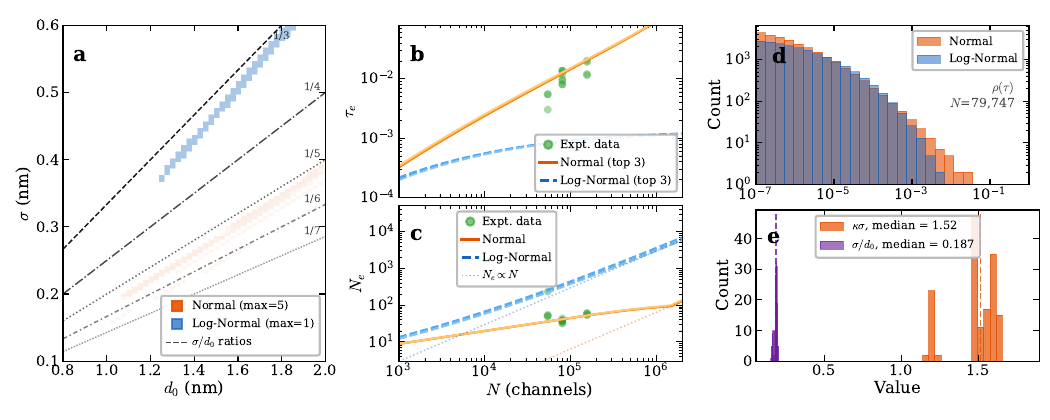}
    \caption{Comparing the area dependence of junction observables assuming normal $\rho(d)\sim\mathcal{N}(d_\mathrm{I},\sigma_\mathrm{I})$ or log-normal $\rho(d)\sim\mathcal{LN}(d_\mathrm{I},\sigma_\mathrm{I})$ oxide thickness distributions.
    \textbf{a},~Colored ($\sigma,d_0$) points denote where, for $0.5<\theta<2.5$~eV, a solution curve passes within $10\%$ of \textit{any} experimental $(N_e,\tau_e)$ pair. Each curve is scored against the 16 dies at the three smallest junction sizes ($A_\mathrm{jj}=0.054$, 0.080 and 0.153~$\mu\mathrm{m}^2$). From a total simulation pool of 288{,}684 valid $(\sigma,d_0,\theta)$ combinations, for $\rho(d)\sim\mathcal{LN}(d_\mathrm{I},\sigma_\mathrm{I})$, 329 curves matched a single die and none matched more than one. For $\rho(d)\sim\mathcal{N}(d_\mathrm{I},\sigma_\mathrm{I})$, 749 curves matched, up to five dies at once, and these curves captured the area dependence much more accurately for $N_e,\tau_e$, as shown in \textbf{b} and \textbf{c}. The dashed lines in \textbf{c} represent the expected $N_e\propto N_\mathrm{ch}$ when $M_1,M_2$ reach convergence.
    \textbf{d},~Example histogram comparing optimum fit curves from (b,c). The normal distribution results in a much wider fat-tailed $\rho(\tau)$. \textbf{e},~The linear dependence in (a) shows that $\theta$ is simply a rescaling parameter for the $(d_0,\sigma)$ solution space. The true metric determining $N_e,\tau_e$ is $\sigma/d_0$. Likewise, every solution has $\omega = \kappa\sigma >1$, well above the fat-tail onset $\omega=0.5$ and the size-dependence threshold $\omega\gtrsim0.8$, so the entire solution space is in the non-self-averaging regime.
    }
    \label{fig:FigS7}
\end{figure}

Although the area-dependent trend of $N_e,\tau_e$ is directly predicted for $\rho(d)$ normal, the question arises as to whether \textit{any} physically reasonable combination of $\sigma,d_0,\theta$ satisfies the actual experimental $\tau_e(A_\mathrm{jj}),N_e(A_\mathrm{jj})$ trends. As a demonstration, we consider the three smallest junction sizes ($A_\mathrm{jj}\le0.153~\mu\text{m}^2$) because $\tau_e$ increases monotonically in this region. OS is then used to calculate the deterministic solution space of normal and log-normal distributions for $d_0\in[0.8~\text{nm},2~\text{nm}]$, $\theta\in [0.5~\text{eV},2.5~\text{eV}]$, and $\sigma\in [0.1~\text{nm},0.6~\text{nm}]$. For a given solution, if $|N_{e,\text{sim}}-N_{e,\text{exp}}|/N_{e,\text{exp}}<0.1$ and $|\tau_{e,\text{sim}}-\tau_{e,\text{exp}}|/\tau_{e,\text{exp}}<0.1$, that parameter set scores $+1$. For every die where this holds we add another point; the 16 dies at the three smallest sizes ($A_\mathrm{jj}=0.054$, 0.080 and 0.153~$\mu\mathrm{m}^2$) are scored. Results for both normal and log-normal $\rho(d)$ are shown in Fig.~\ref{fig:FigS7}a, with the solution spaces resulting from different $(d_0,\sigma,\theta)$ values. Curves sit along constant $\sigma/d_0\sim1/5$ (normal) and $\sigma/d_0\sim1/3$ (log-normal). The collinearity of the points indicates that $\theta$ acts only as a scaling parameter within the model, and can be chosen based on TEM data or literature values for the oxide and contacts. As Fig.~\ref{fig:FigS7}b,c shows, \textit{no} combinations from the log-normal distribution fit more than a single experimental $(N_e,\tau_e)$ data point. The rapid convergence of $M_1,M_2$ results in (i) a linear $N_e$ vs.\ $A_\mathrm{jj}$ relationship instead of the sublinear dependence observed experimentally for small areas, and (ii) $\tau_e\approx\text{const}$, instead of the linear increase. These results are consistent with the convergence analysis presented in Fig.~\ref{fig:FigS5} and Fig.~\ref{fig:FigS6}, but shown explicitly to hold also for the experimental data. We therefore conclude that the underlying $\rho(\tau)$ is described best by a normal distribution (Eq.~\ref{eq:S6}). We cannot rule out the possibility that the normal distribution skews right, but this upper tail is irrelevant to the tunneling transparencies and adds unnecessary complexity. This result reflects the properties of the lower tail of the oxide thickness distribution, where the majority of tunnel current flows.

For a normal distribution, convergence of $M_2$ to the true average requires $N_\mathrm{ch}\gg10^6$ or $A_\mathrm{jj}>1~\mu\text{m}^2$, which is larger than standard junction sizes in quantum computing circuits. The histograms of the $\tau$ distributions from (b),(c) are shown in Fig.~\ref{fig:FigS7}d on a log scale. As predicted above, fewer than $\sim 10$ channels in the fat tail account for the differences between normal and log-normal. As Fig.~\ref{fig:FigS7}e shows, the solution space collapses onto lines of constant $\sigma/d_0$: because $\theta$ enters only through $\kappa$, $N_e$ and $\tau_e$ depend on $(d_0,\sigma,\theta)$ through $\kappa d_0$ and $\omega=\kappa\sigma$ alone, and $\sigma/d_0=\omega/(\kappa d_0)$ is the $\theta$-independent invariant of the fit. Every solution has $\omega>1$, which places the entire space beyond both thresholds established above: the fat-tail onset $\omega=0.5$ (Eq.~\ref{eq:S11}), above which $M_2$ is sourced by a rare thin-channel tail rather than by channels near $d_0$, and the size-dependence threshold $\kappa\sigma\gtrsim0.8$ quoted in the main text, above which the sampled $\tau_e$ varies over the experimental window $10^4<N_\mathrm{ch}<2\times10^6$ (strongly so for $\kappa\sigma\gtrsim1$). The Region-I fits of Table~\ref{tab:TableS2} ($\omega_\mathrm{I}=1.2$ to $1.6$) lie in the same regime. Below the onset, by contrast, all accessible moments are carried by channels within $\sim\sigma$ of $d_0$ and the transparency statistics self-average, so $\tau_e$ is size-independent and $N_e\propto N_\mathrm{ch}$ (main text, Fig.~\ref{fig:Fig4}b,c); Region II, with $\omega_\mathrm{II}\approx0.55$ (Sec.~\ref{sec:S5}), sits below the size-dependence threshold and self-averages over the measured range. $\omega$ is therefore the figure of merit for junction quality: it sets both the convergence of $M_k^p$ with $N_\mathrm{ch}$ and the tail enhancement $\xi$ (Fig.~\ref{fig:FigS5}), while $\sigma/d_0$ fixes it for a given barrier height.

The remainder of the analysis will proceed considering only a normal distribution for $\rho(d)$.

\FloatBarrier
\section{Model Fitting}
\label{sec:S5}

\subsection{Small-Junction Fitting}

We fit the small-junction (pure Region-I) regime using:

\begin{equation}
\begin{split}
\mathcal{L}(d_\mathrm{I},\sigma_\mathrm{I})=&\sum_{i}[(\log  \tau_{e,i}^{model}-\log  \tau_{e,i}^{experiment})^2\\
&+~ (\log  N_{e,i}^{model}-\log  N_{e,i}^{experiment})^2\\
&+~ (\log  (\dfrac{G_n}{A_\mathrm{jj}})_{i}^{model}-\log  (\dfrac{G_n}{A_\mathrm{jj}})_{i}^{experiment})^2]
\end{split}
\label{eq:S31}
\end{equation}

where $\mathcal{L}(d_\mathrm{I},\sigma_\mathrm{I})$ is the cost function. As shown in Fig.~\ref{fig:FigS7}, the solution space for $(d_\mathrm{I},\sigma_\mathrm{I},\theta_\mathrm{I})$ is not unique, but rather scales with the choice of $\theta$. Further, $\sigma_\mathrm{I}/d_\mathrm{I}\approx0.19$ for all samples tested in this study, which limits the fitting so that optimization is constrained and a global minimum can be found.

\begin{figure}[!tbp]
    \centering
    \includegraphics[width=.5\linewidth]{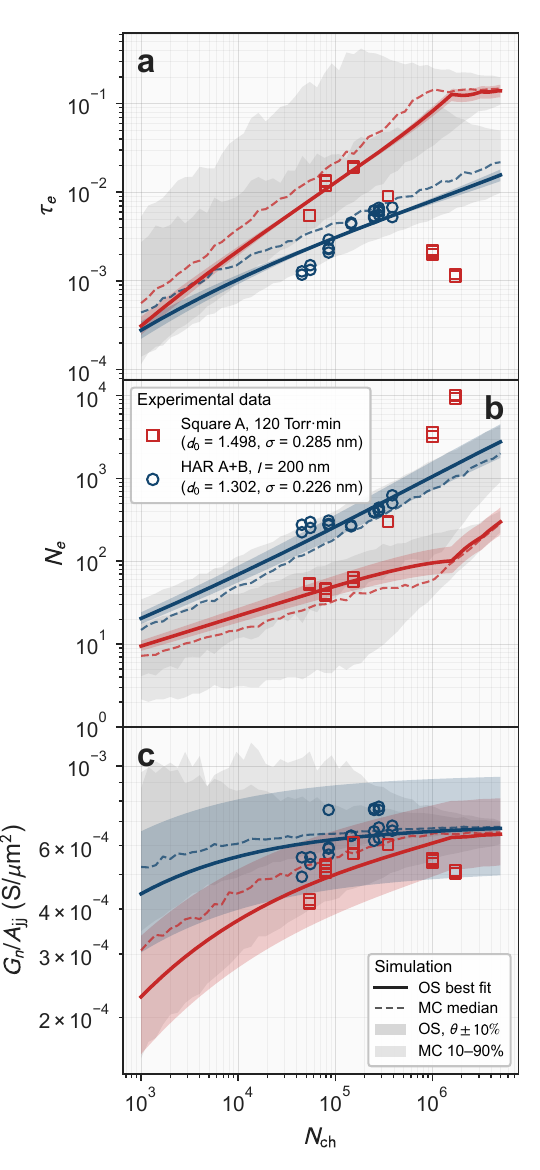}
    \caption{Small-area fits: order statistics (OS) and Monte Carlo simulations. The OS fit is performed on the pooled $l = 200$~nm HAR junctions (Sets A and B) and on the Square Set A 120~Torr$\cdot$min junctions at the two smallest areas ($N_\mathrm{ch}< 10^5$) using Eq.~\ref{eq:S31} with parameters listed in Table~\ref{tab:TableS2}. The $\pm10\%$ error band uses fixed $\sigma_\mathrm{I},d_\mathrm{I}$ while varying $\theta_\mathrm{I}$. MC simulations show 80\% CI using the same parameters extracted from the OS fitting. No correlations are included in the MC simulations, therefore due to the non-self-averaging behavior of $\tau_e,N_e$ the resulting error bars are an order of magnitude larger than experimentally (or `typically') observed. If we re-solve for $\sigma_\mathrm{I},d_\mathrm{I}$ while varying $\theta_\mathrm{I}$, the resulting error bands are invisible behind the original line, further confirming that $\theta_\mathrm{I}$ acts only as a scaling parameter for junctions with the same fabrication process.
    \textbf{a},~$\tau_e$ vs.\ $N_\mathrm{ch}$.
    \textbf{b},~$N_e$ vs.\ $N_\mathrm{ch}$.
    \textbf{c},~$G_n/A_\mathrm{jj}$ vs.\ $N_\mathrm{ch}$ trends are correctly reproduced. The $A_\mathrm{jj}$ dependence of $G_n/A_\mathrm{jj}$ results from a large $\omega_\mathrm{I}=\kappa \sigma_\mathrm{I}$ value which prevents $M_1=G_n h/(2e^2 N_\mathrm{ch})$ from converging for $A_\mathrm{jj}<2~\mu\text{m}^2$. This is a consequence of a fat-tailed distribution, and can be reduced by fabricating junctions with smaller $\sigma_\mathrm{I}/d_\mathrm{I}$. Varying $\theta_\mathrm{I}=1.15\pm0.115$~eV results in $G_n/A_\mathrm{jj}$ changes that are far greater than experimentally observed; therefore this disorder mechanism is not present in the junctions.}
    \label{fig:FigS8}
\end{figure}

Results are shown in Fig.~\ref{fig:FigS8}. The OS method of extracting parameters is also confirmed using the same $d_\mathrm{I},\sigma_\mathrm{I},\theta_\mathrm{I}$ in MC simulations (1000 runs). MC has the advantage of showing the entire solution space, compared to OS which gives the typical result. Unfortunately, in non-self-averaging systems the average of the entire solution space implies randomness 1--2 orders of magnitude larger than experimentally observed. For this reason, channel homogeneity is handled more deterministically with order statistics. As shown in  Fig.~\ref{fig:FigS8}, the MC median value tracks with the OS value with the same input parameters for $d_\mathrm{I},\theta_\mathrm{I},\sigma_\mathrm{I}$. Large statistics of devices (hundreds), which is not experimentally viable, would be needed to account for the rare extreme events.

\subsection{Region-II and Multi-device Fitting}

$\theta_\mathrm{I} = \theta_\mathrm{II} = 1.15$~eV and the Region-II roughness $\sigma_\mathrm{II} = 0.10$~nm are held fixed, with $\theta$ serving as a scaling factor during combined cost minimization to find the remaining free parameters: $d_\mathrm{I}(\mathcal{E})$, $d_\mathrm{II}(\mathcal{E})$, and the threshold length $l_\mathrm{th}$ that caps the Region-I area at $A_\mathrm{th} = l_\mathrm{th}^2 + 2\,l_\mathrm{th}\,t_\mathrm{b}$ once $\min(l,w)\ge l_\mathrm{th}$ (main text, Eq.~\ref{eq:4}). Results of the fitting are shown in Fig.~\ref{fig:Fig4} in the main text. These results are deterministic and identical for all junctions of a given area $A_\mathrm{jj}$.

\begin{figure}[!tbp]
    \centering
    \includegraphics[width=1\linewidth]{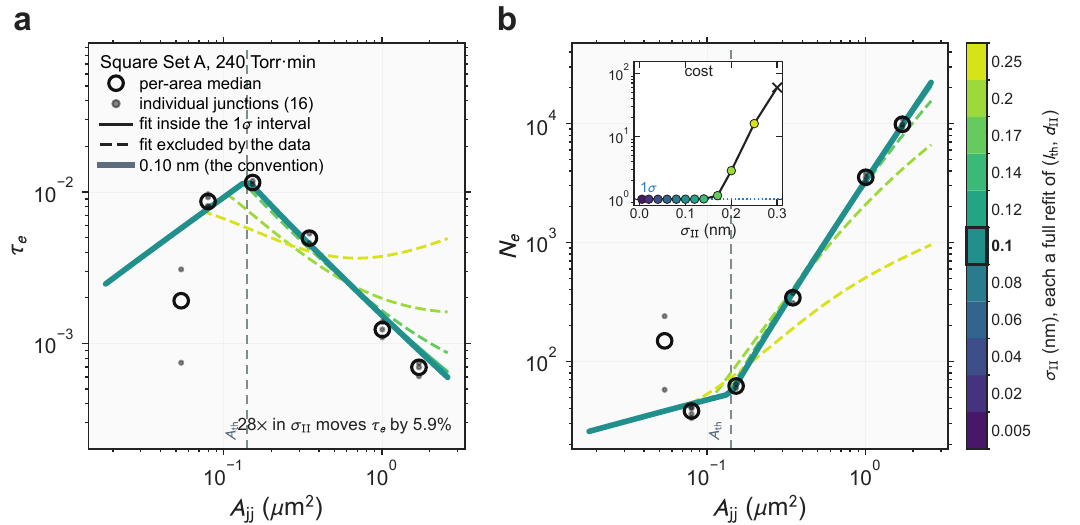}
    \caption{Bounding the Region-II roughness $\sigma_\mathrm{II}$: the two-region joint fit for the 240~Torr$\cdot$min Square Set A condition (16 junctions over six areas; open circles are the per-area medians the cost is scored on, small dots individual junctions), re-solved with $\sigma_\mathrm{II}$ pinned at twelve values from 0.005 to 0.30~nm (colorbar), refitting $l_\mathrm{th}$ and $d_\mathrm{II}(\mathcal{E})$ at each.
    \textbf{a},~$\tau_e$ versus $A_\mathrm{jj}$. Solid curves lie inside the $1\sigma$ cost interval ($\sigma_\mathrm{II}\le0.14$~nm) and form a single indistinguishable bundle: a $28\times$ range of $\sigma_\mathrm{II}$, compensated by $\lesssim60$~pm of $d_\mathrm{II}$, moves $\tau_e$ by only 6\%. Dashed curves are excluded by the data. The vertical line marks $A_\mathrm{th}$; below it the junction is pure Region I and $\sigma_\mathrm{II}$ has no channels to describe.
    \textbf{b},~$N_e$ versus $A_\mathrm{jj}$, same encoding. Inset: the fit cost versus $\sigma_\mathrm{II}$, flat to 1\% over 0.005--0.14~nm and then rising steeply ($3\times$ at 0.20~nm, $60\times$ at 0.30~nm, marked $\times$); the dotted line is the $1\sigma$ threshold. The transport data therefore set only an upper bound on $\sigma_\mathrm{II}$; the convention $\sigma_\mathrm{II}=0.10$~nm (thick curve; boxed value on the colorbar) lies inside the allowed plateau.}
    \label{fig:FigS9}
\end{figure}

Holding $\sigma_\mathrm{II}$ fixed is a convention and not a fit result. The transport data bound $\sigma_\mathrm{II}$ from above but the fit does not resolve it. Re-solving the joint fit with $\sigma_\mathrm{II}$ pinned at values spanning 0.005--0.30~nm, refitting $l_\mathrm{th}$ and $d_\mathrm{II}(\mathcal{E})$ at each (Fig.~\ref{fig:FigS9}), leaves the cost flat to 1\% up to $\sigma_\mathrm{II}=0.14$~nm, giving a $1\sigma$ upper bound $\sigma_\mathrm{II}\lesssim0.14$~nm with no lower limit. Within this plateau $\sigma_\mathrm{II}$ is degenerate with $d_\mathrm{II}$: both act only through the thin-barrier tail of $\tau(d)$, and a $28\times$ change in $\sigma_\mathrm{II}$ is absorbed by a $\lesssim60$~pm shift of $d_\mathrm{II}$. The upper bound is algebraic, since Region II must remain self-averaging over the measured range. Therefore $\omega_\mathrm{II}=\kappa\sigma_\mathrm{II}\lesssim0.8$ or \ $\sigma_\mathrm{II}\lesssim0.15$~nm at $\theta=1.15$~eV. Beyond this threshold the Region-II tail becomes non-self-averaging and dominates $I_2$, erasing the small/large-junction contrast that defines $l_\mathrm{th}$. As $\sigma_\mathrm{II}$ rises, the fit cost rises accordingly by $3\times$ at $\sigma_\mathrm{II}=0.20$~nm ($\omega_\mathrm{II}=1.1$) and $60\times$ at $0.30$~nm. The adopted $\sigma_\mathrm{II}=0.10$~nm ($\omega_\mathrm{II}=0.55$) sits inside the allowed plateau, well below the fitted Region-I roughness $\sigma_\mathrm{I}\approx0.28$--$0.29$~nm; the residual conventional freedom enters the quoted $d_\mathrm{II}$ only as a $\pm60$~pm systematic.

\FloatBarrier

\end{document}